\documentclass[12pt]{article}
\pdfoutput=1

\usepackage{amsmath}
\usepackage{amsthm}
\usepackage{amscd}
\usepackage{amsfonts}
\usepackage[psamsfonts]{amssymb}
\usepackage{graphicx}
\usepackage{epsfig}
\usepackage{hyperref}
\usepackage{color}

\makeatletter
\renewcommand\section{\@startsection {section}{1}{\z@}%
                  {-3.5ex \@plus -1ex \@minus -.2ex}
                  {2.3ex \@plus.2ex}%
                  {\normalfont\large\bfseries}}
\renewcommand\subsection{\@startsection{subsection}{2}{\z@}%
                  {-3.25ex\@plus -1ex \@minus -.2ex}%
                  {1.5ex \@plus .2ex}%
                  {\normalfont\bfseries}}
\renewcommand\subsubsection{\@startsection{subsubsection}{3}{\z@}%
                  {-3.25ex\@plus -1ex \@minus -.2ex}%
                  {1.5ex \@plus .2ex}%
                  {\normalfont\itshape}}
\makeatother

\def\baselinestretch{1.2}
\begin{document}

\begin{titlepage}
\begin{flushright}
CQUeST-2013-0593 \\
TAUP-2963/13
\end{flushright}
\vskip1cm

\begin{center}
{\Large{\bf Aging Logarithmic Galilean Field Theories}}

\bigskip\bigskip
Seungjoon Hyun$^{1}$, Jaehoon Jeong$^{2}$ and
Bom Soo Kim$^{3}$
\bigskip

${}^1${\it {\small Department of Physics, College of Science, Yonsei University, Seoul 120-749, Korea}} \\
${}^2${\it {\small Center for Quantum Spacetime, Sogang University, Seoul 121-742, Korea}}\\
${}^3${\it {\small Raymond and Beverly Sackler School of Physics and Astronomy,}} \\
{\it {\small Tel Aviv University, 69978, Tel Aviv, Israel}}\\

\end{center}

\bigskip 

{\small
\centerline{sjhyun@yonsei.ac.kr, ~~jhjeong@sogang.ac.kr, ~~bskim@post.tau.ac.il}
}

\bigskip\bigskip
\begin{abstract} 

We analytically compute correlation and response functions of scalar operators for the
systems with Galilean and corresponding aging symmetries for general 
spatial dimensions $d$ and dynamical exponent $z$, along with their logarithmic 
and logarithmic squared extensions, using the gauge/gravity duality. 
These non-conformal extensions of the aging geometry are marked by 
two dimensionful parameters, eigenvalue $\mathcal M$ of an internal coordinate 
and aging parameter $\alpha$. 

We further perform systematic investigations on two-time response functions for general $d$ and $z$, 
and identify the {\it growth exponent} as a function of the scaling dimensions $\Delta$ of 
the dual field theory operators and aging parameter $\alpha$ in our theory. 
The initial growth exponent is only controlled by $\Delta$, 
while its late time behavior by $\alpha$ as well as $\Delta$. 
These behaviors are separated by a time scale order of the waiting time.
We attempt to make contact our results with some field theoretical growth models, 
such as Kim-Kosterlitz model at higher number of spatial dimensions $d$.   

\end{abstract}

\vspace{0.5in}
\end{titlepage}

\def\baselinestretch{1.0}
\tableofcontents

\def\baselinestretch{1.2}

\section{Introduction}

Non-equilibrium growth and aging phenomena are of great interest due to their wide applications 
across various scientific fields of study, including many body statistical systems, 
condensed matter systems, biological systems and so on 
\cite{HenkelBook1}-\cite{HinrichsenReviewDP}.
They are complex physical systems, and details of microscopic dynamics are 
widely unknown. Thus it is best to describe these systems with a small number of variables, 
their underlying symmetries and corresponding universality classes, which have been 
focus of nonequilibrium critical phenomena. 

One particular interesting class is described by the Kardar-Parisi-Zhang (KPZ) equation 
\cite{KPZ}\cite{BSBook}. Recently, this class is realized in a clean experimental setup 
\cite{TakeuchiPRL}\cite{TakeuchiSciRep}, and their exponents for one spatial dimension 
$d=1$ is confirmed: the roughness $\mathrm a = \frac{1}{2} $, 
the growth $\mathrm b = \frac{1}{3} $ and the dynamical $z = \frac{3}{2}$ exponents.  
Along with the experimental developments, there have been also theoretical developments 
in the context of aging. The KPZ class reveals also simple aging in the two-time 
response functions \cite{Henkel:2011NP}\cite{Henkel:2010hd}. 
In these works, it is shown that the autoresponse function of the class is well described 
by the logarithmic (log) and logarithmic squared (log$^2$) extensions of the scaling function 
with local scale invariance for $d=1$. 

In a recent paper \cite{Hyun:2012fd} based on \cite{Jottar:2010vp}\cite{Hyun:2011qj}, 
we have considered log extensions of 
the two-time correlation and response functions of the scalar operators with the conformal 
Schr\"odinger and aging symmetries for the spatial dimension $d=2$ 
and the dynamical exponent $z=2$, in the context of gauge/gravity duality 
\cite{Maldacena:1997re}\cite{Aharony:1999ti}. 
The power-law and log parts are determined by the scaling dimensions of 
the dual field theory operators, the eigenvalue of the internal coordinate 
and the aging parameter, which are explained below. 
Interestingly, our two-time response functions show several
qualitatively different behaviors: growth, aging (power-law decaying) or both behaviors 
for the entire range of our scaling time, depending on the parameters in our theory 
\cite{Hyun:2012fd}. 

We further have made connections to the phenomenological field theory model 
\cite{Henkel:2011NP} in detail. 
The two-time response functions and their log corrections of our holographic model 
\cite{Hyun:2012fd} are completely fixed by a few parameters and are valid for 
$z=2, d=2$, while the field theory model \cite{Henkel:2011NP} has log$^2$ extensions 
and is valid for $z=\frac{3}{2}, d=1$. Closing the gap between these two models from 
the holographic side is one of the main motivation of this work.  

In this paper, we generalize our analysis \cite{Hyun:2012fd} in two different 
directions: (1) by applying to general dynamical exponent $z$ (not conformal) as well as 
to any number of spatial dimensions $d$ and 
(2) by including log$^2$ corrections in two-time response functions. 
In \S \ref{sec:NTLFT}, we first analytically compute the correlation and response functions for 
general $z$ and $d$ along with their log and log$^2$ extensions.  
Then we add the aging generalizations of our non-conformal results in \S \ref{sec:Aging}. 
We try to make contact with KPZ class in \S \ref{sec:KPZ}, and conclude in \S \ref{sec:conclusion}.

\section{Logarithmic Galilean Field Theories} \label{sec:NTLFT} 

Logarithmic conformal field theory(LCFT) is a conformal field theory(CFT) which contains correlation functions with logarithmic divergences.\footnote{ See, for example, \cite{Flohr:2001zs}\cite{Gaberdiel:2001tr} for the reviews on LCFT.} They typically appear when two primary operators with the same conformal dimensions are indecomposable and form a Jordan cell.
The natural candidates for the bulk fields, in the holographic dual descriptions of LCFT, of the pair of two primary operators forming Jordan cell are given by a pair of fields with the same spin and a special coupling. After integrating out one of them, it becomes the model with higher derivative terms. 

The AdS dual construction of the LCFT was first considered in \cite{Ghezelbash:1998rj}\cite{Kogan:1999bn}\cite{Myung:1999nd} using a higher derivative scalar field on AdS background. Recently, higher derivative gravity models on AdS geometry with dual LCFT  have got much attention, starting in three dimensional gravity models \cite{Grumiller:2008qz}-\cite{Alishahiha:2010bw}. In four and higher dimensional AdS geometry, the gravity models with curvature-squared terms typically contain massless and ghostlike massive spin two fields.\footnote{By imposing an appropriate boundary condition for the ghost-like massive mode which falls off more slowly than the massless one, it was argued in \cite{Maldacena:2011mk}\cite{Hyun:2011ej}\cite{Hyun:2012mh} that  the theory after the truncation becomes
effectively  the usual Einstein gravity at the classical but nonlinear level.} When the couplings of the curvature-squared terms are tuned, the massless and massive modes become degenerate and turn into the massless
and logarithmic modes \cite{Lu:2011zk}. This, so called, critical gravity has the boundary dual LCFT which contains stress-energy tensor operator
and its logarithmic counterpart.

More recently, studies on generalizations of LCFT in the context of AdS/CFT correspondence \cite{Maldacena:1997re}\cite{Aharony:1999ti}, in particular the correlation functions of a pair of scalar operators, have been made in two different directions. One is on the non-relativistic LCFT. In \cite{Bergshoeff:2011xy} the dual LCFT to the scalar field  in the Lifshitz background has been investigated. The study on the dual LCFT to the scalar field in the Schr\"odinger and Aging background was made in \cite{Hyun:2011qj}. The other is on the LCFT with $\log^2$ divergences. Correlation functions with log$^2$ corrections have been investigated in several works. In the context of the gravity modes of tricritical point, they are interpreted as 
rank-3 logarithmic conformal field theories (LCFT) 
with log$^2$ boundary conditions \cite{Bergshoeff:2012ev}. 
Explicit action for the rank-3 LCFT is considered in \cite{Moon:2012vc}. 
See also a recent review on these developments in \cite{Grumiller:2013at}. 

In this section we would like to accomplish two different things motivated by these developments 
along with those explained in the introduction. 
First, we compute correlation and response functions for AdS in light-cone (ALCF) 
and Schr\"odinger type gravity theories, which are dual to some non-relativistic field theories 
with Galilean invariance for general dynamical exponent $z$ and $d$. 

Second, we generalize our correlation and response 
functions with log$^2$ as well as log contributions 
for $z$ and $d$. 
The log correction has been investigated in \cite{Hyun:2012fd} in the context of Schr\"odinger 
geometry and Aging geometry for $d=2$ and $z=2$ as mentioned in the introduction. 

We first consider the ALCF \cite{Goldberger:2008vg}\cite{Barbon:2008bg}. 
Due to several technical differences,  
we present our computations in some detail, building up correlation functions \S \ref{sec:ALCFCorr}, 
their log extensions \S \ref{sec:ALCFCorrLog} and log$^2$ extension \S \ref{sec:ALCFCorrLog2}. 
For Schr\"odinger backgrounds \cite{Son:2008ye}\cite{Balasubramanian:2008dm}, 
we comment crucial differences in 
\S \ref{sec:SchrLog}. And then we present the correlation functions for $z=2$ in \S \ref{sec:zTwo} 
and for $z=3/2$ in \S \ref{sec:zThreeHalf}. 
We summarize our results in \S \ref{sec:NTLFTsummary}.

\subsection{ALCF}  \label{sec:ALCF}

Let us turn to the AdS in light-cone (ALCF) with Galilean symmetry studied in 
\cite{Goldberger:2008vg}\cite{Barbon:2008bg}\cite{Kim:2012nb}. The case for general $ z $ is 
also considered in \cite{Kim:2012nb} for zero temperature and in \cite{Kim:2012pd} 
for finite temperature. The metric is given by 
	\begin{align}  \label{ALCFMetric}
		ds^2 = L^2  \frac{d\vec x^2 - 2 dt d\xi + du^2}{u^{2}}  \;,  
	\end{align}
which is invariant under the space-time translations $P_i, H$, Galilean boost $K_i$, 
	\begin{align}  \label{GalileanTR1}
		\vec x \to \vec x - \vec v t \;,  \qquad \xi \to \xi 
		+ \frac{1}{2} (2 \vec v \cdot \vec x - v^2 t) \;, 
	\end{align}
scale transformation $ D $,
	\begin{align}  \label{ScaleTR1}
		t \to \lambda^z t \;, \quad \vec x \to \lambda \vec x \;, \quad 
		u \to \lambda u \;, \quad \xi \to \lambda^{2-z} \xi \;, 
	\end{align}
and translation along the $ \xi $ coordinate, which represents the dual particle number or rest mass.%
\footnote{
Apparently there exists another symmetry transformation, which is the following special conformal 
one for general $z$, 
	\begin{align}   \label{ALCFScaling}
		t \to \frac{t}{1+c t} \;, \quad \vec x \to \frac{\vec x}{1+c t} \;, \quad 
		u \to \frac{u}{1+c t} \;, \quad 
		\xi \to \xi + \frac{c}{2} \frac{\vec x \cdot \vec x + r^2}{1+c t} \;.
	\end{align}
It turns out that this does not provide a closed algebra with other symmetry generators 
for $z \neq 2$.} 
The geometry satisfies vacuum Einstein equations with a negative cosmological constant.
The finite temperature generalizations of the ALCF for $z=2$ and $d=2$ have been considered in 
\cite{Maldacena:2008wh}\cite{Kim:2010tf}\cite{Kim:2010zq}.

\subsubsection{Correlation functions}   \label{sec:ALCFCorr}

We compute correlation functions of the geometry (\ref{ALCFMetric}) by 
coupling a probe scalar.  
\begin{align}   \label{ScalarActionALCF}
	S = K \int d^{d+2} x \int_{u_B}^{\infty} du \sqrt{-g}
	& \left( \partial^M \phi^* \partial_M \phi + m^2 ~ \phi^* \phi \right)  \;, 
\end{align}
where $K $ is a coupling constant, and $M, N = u, t, \xi, \vec x$.
We use $u_B$ for our boundary cutoff. 
The field equation of $\phi$ for general $z$ and $d$ is
\begin{align}
 \frac{\partial^2 \phi}{\partial u^2}  -(d+1) \frac{1}{u} \frac{\partial \phi}{\partial u} 
-\left ( \frac{ m^2 L^2 }{u^2} +\vec k^2 +2 M w  \right) \phi = 0
  \;.
\label{BlukScalarEqALCF}
\end{align}
Note that we treat $\xi$ coordinate special and replace all the $\partial_{\xi}$ as $iM$. 
This is in accord with the fact that the coordinate $\xi$ plays a distinguished role in 
the geometric realizations of Schr\"odinger and Galilean symmetries  
\cite{Son:2008ye}\cite{Balasubramanian:2008dm}\cite{Kim:2012nb}. 

The general solution is given by 
\begin{align}
	\phi = c_1 u^{1+d/2}  I_\nu (q u) + c_2 u^{1+d/2} K_\nu (q u ) \;, 
\end{align}
where $\nu = \pm \sqrt{(1+d/2)^2+L^2 m^2}, q^2 = \vec k^2 + 2M\omega$, $I, K$ represent Bessel functions. 
We choose $K$ over $I$ due to its well defined properties deep in the bulk.  

We follow \cite{Son:2002sd}\cite{Hyun:2011qj} to compute correlation functions 
by introducing a cutoff $u_B$ 
near the boundary and normalizing $f_{\omega,\vec{k}}(u_B)  = 1$, 
which fixes $ c =  u_B^{-1-d/2} K_\nu^{-1} (q u_B)$. 
We compute an on-shell action to find  
\begin{align}  \label{OnShAct1}
	S[\phi_0] 
	&= \int d^{d+1} x  \frac{L^{d+3}}{u^{d+3}}  ~\phi^* (u,t, \vec x) ~\frac{u^2}{L^2}\partial_u \phi (u,t, \vec x) \big |_{u_B} \;. 
\end{align}
Using  $k = (\omega, \vec k)$, $x = (t, \vec x) $, 
$i k \cdot x = -i \omega t + i \vec{k} \cdot \vec{x} $ and
\begin{align}
\phi(u, x) = \int \frac{d \omega}{2 \pi} \frac{d^d k}{(2\pi )^d} ~e^{i k \cdot x} 
~ \phi_{k}(u) ~\phi_0 (k)  \;,
\label{IntTran}
\end{align}
the onshell action can be rewritten as 
\begin{align}	
	& \int [t, \omega', \omega] \int [x, \vec k', \vec k] \quad  \phi_0^* (k') \mathcal F_1 (u,k',k) \phi_0 (k) \Big |_{u_B} \;, 
	\label{wholeEQ}
\end{align}
where $ \int [t, \omega', \omega] \int [x, \vec k', \vec k]  
= \int d t ~\frac{d \omega'}{2 \pi} \frac{d \omega}{2 \pi} e^{-i (\omega' -\omega) t} 	
	\int d^d x  \int \frac{d^d k'}{(2\pi)^d} \frac{d^d k}{(2\pi)^d} 
	e^{i(\vec k' - \vec k) \cdot \vec{x}}  $.  
$\mathcal F_1$ is given by 
\begin{align}   
	\mathcal F_1 (u, k', k)
	&= \frac{L^{d+3}}{u^{d+3}} \phi_{k'}^* (k',u) 
	\frac{u^2}{L^2}\partial_u \phi_{k} (k,u) \;.
\end{align}
This function appears again when we construct the log and log$^2$ extensions below. 

For general $d$ we find 
\begin{align}
\frac{\partial}{\partial u}  u^{1+d/2}   K_\nu (q u) 
= u^{d/2} \{ (1+d/2-\nu)  K_\nu (q u) - q u K_{\nu-1}  (q u) \} \;.
\end{align}
One can evaluate the $q$-dependent part of $\mathcal F_1$ at the boundary, $u=u_B$, by 
expanding it for small $u_B$. We obtain the following non-trivial contribution 
\begin{align}    \label{F1Function}
	\mathcal F_1 (u_{B},\omega,\vec{k})
	&=  - \frac{2 \Gamma (1-\nu)}{ \Gamma (\nu)}
	\left( \frac{L^{d+1}}{u_B^{d+2}} \right)    \left( \frac{q u_B}{2}\right)^{2\nu} + \cdots   \;.
\end{align}
Note that the function $\mathcal F_1$ is only a function of $q (\omega, \vec k)$ 
when it is evaluated at the boundary. 

By inverse Fourier transform of (\ref{F1Function}) 
for an imaginary parameter $M = i \mathcal M$, we get the following coordinate space 
correlation functions for a dual field theory operator $\phi$
\begin{align}
\langle \phi^* (x_{2}) \phi (x_{1}) \rangle_{\mathcal F_1} &=
-2 \theta (t_2)~ \int  \frac{d \omega}{2 \pi} \frac{d^d k}{(2\pi )^d} 
 e^{-i \vec{k} \cdot (\vec{x}_{1}-\vec{x}_{2})} e^{ i \omega (t_{1} - t_2) }  
 \mathcal F_1 (u_{B},k)   \label{F1Correlator} \\
&= \frac{\Gamma (1-\nu)}{ \Gamma (\nu) \Gamma (-\nu)}  
\frac{L^{d+1} \mathcal M^{\frac{d}{2}+\nu}  }{\pi 2^{\nu-1}  u_B^{d+2-2\nu}} 
\cdot \frac{\theta (t_2) \theta(t_2 - t_1) }{ (t_2 - t_1)^{\frac{d+2}{2}+\nu}}   
\cdot \exp \left( -{ \frac{\mathcal M (\vec{x}_2 - \vec{x}_1)^2 }{2 ( t_2 - t_1)}}\right)  \;. 
\nonumber
\end{align}
These are our correlation functions evaluated for general dynamical exponent $z$ and 
for the number of spatial dimensions $d$, which are direct 
generalizations of a previous result for $d=2$ and $z=2$ \cite{Hyun:2011qj}. 
Note that the result is valid for the field theories with Galilean boost without 
conformal symmetry. In particular, the parameter $\mathcal M$ carries scaling dimensions 
$[\mathcal M] = z-2$, and the exponent is actually dimensionless. This is consistent 
with the scaling properties written in (\ref{ALCFScaling}). 

The result (\ref{F1Correlator}) is independent of $z$, while depending on the number of dimensions $d$.  
This is expected because the ALCF metric (\ref{ALCFMetric}) is independent of $z$, 
which is a special feature for the ALCF. 
This is not true for Schr\"odinger background as we see below.

\subsubsection{Response functions with log extension}   \label{sec:ALCFCorrLog}

Motivated by the recent interests on LCFT from the holographic point of view 
\cite{Bergshoeff:2011xy}\cite{Hyun:2012fd}, 
we consider two scalar fields $\phi$ and $\tilde \phi$ in the background (\ref{ALCFMetric}) 
	\begin{align}  \label{SchrActionLCFT}
		S = K \int d^{d+2} x \int_{u_B}^{\infty} du \sqrt{-g}
		& \left(  \partial^M \tilde \phi ~\partial_M \phi
		+ m^2 ~\tilde  \phi ~\phi +\frac{1}{2L^2}\tilde \phi^2 \right)  \;,
	\end{align}
where $u_B $ represents a cutoff near the boundary. We take $\partial_\xi \phi=-{\cal M}\phi$ and
$\partial_\xi \tilde\phi=- {\mathcal M}\tilde\phi$. 

The field equations for $\phi$ and $\tilde \phi$ of the action (\ref{SchrActionLCFT}) become
	\begin{align}   \label{DiffEqSchr}
		&2{\cal M} \partial_t \phi =
		- \frac{1}{u^2} \mathcal D  \phi - \vec \nabla ^2  \phi + \frac{1}{u^2} \tilde \phi \;, \\
		&2{\cal M} \partial_t \tilde \phi =
		- \frac{1}{u^2} \mathcal D \tilde \phi - \vec \nabla ^2 \tilde \phi  \;,
	\end{align}
where
	\begin{align} \label{DiffOpALCF}
		\mathcal  D = u^2 \bigg[ \partial_u^2 - \frac{d+1}{u} \partial_u
		- \frac{ m^2 L^2 }{u^2} \bigg] \;,
	\end{align}
which is a differential operator for ALCF.

Following \cite{Hyun:2011qj}\cite{Bergshoeff:2011xy},
we construct bulk to boundary Green's functions $G_{ij} (u, \omega, \vec{k})$ 
	\begin{align}   \label{BulkToBoundaryGreensFunctionLCFT}
		&\phi (u,x) = \int \frac{d^{d+1} k}{(2\pi )^{d+1}}  e^{i k \cdot x }
		\left[ G_{11} (u, k) J(k)+  G_{12} (u, k)\tilde J(k)\right]\;, \nonumber  \\
		&\tilde  \phi (u,x) = \int \frac{d^{d+1} k'}{(2\pi )^{d+1}} e^{-i k' \cdot x }
		\left[  G_{21} (u, k') J(k')+ G_{22} (u, k') \tilde J(k')\right]\;.
	\end{align}
We have $G_{21}=0$, which follows from the structure of the equations of motion given in (\ref{DiffEqSchr}) and the action (\ref{SchrActionLCFT}). The Green's functions satisfy
	\begin{align}   \label{3DifferentialEquations}
	\mathrm D G_{11} &= 0  \;, \qquad
	\mathrm  D G_{12} = G_{22}   \;, \qquad
	\mathrm  D G_{22} = 0   \;,
	\end{align}
where $\mathrm D = \mathcal D- q^2 u^2$ and $q = \sqrt{\vec k^2 +2 {\cal M} i\omega}$.
Solutions of $G_{11}$ and $G_{22}$ are given by
	\begin{align}		
		&G_{11}(u, k_\mu)=  c_{11}~u^{1+d/2} K_\nu (q u)\;, \qquad
		G_{22}(u, k_\mu)=  c_{22}~u^{1+d/2} K_\nu (q u) \;,
	\end{align}
where $\nu = \pm \sqrt{(1+d/2)^2 + m^2L^2} $.
The normalization constants $c_{11}=c_{22}= (u_B^{1+d/2} K_\nu (q u_B))^{-1} 
$ can be determined
by requiring that $G_{11}(u_B, k_\mu) =G_{22}(u_B, k_\mu) =1$ \cite{Son:2002sd}\cite{Hyun:2011qj}.

There exists another Green's function $G_{12}$ due to a coupling between $\phi$ and 
$\tilde \phi$ in the action (\ref{SchrActionLCFT}), which satisfies
	\begin{align}
	\mathrm D G_{12} = G_{22}   \;.
	\end{align}
To evaluate $G_{12}$, we use the same methods used in \cite{Kogan:1999bn}\cite{Hyun:2012fd}.
Using
    \begin{align}   \label{CommutationRelation}
        \left[ \mathrm  D \;, \frac{d}{d \nu} \right]  = 2 \nu \;,
    \end{align}
and the fact that $ \mathrm  D G_{22}=0 $, we get
    \begin{align}
        \mathrm  D \left( \frac{1}{2 \nu} \frac{d}{d\nu} G_{22} \right) =G_{22} \;.
    \end{align}
Thus we have an explicit form.  
	\begin{align}   \label{G12ExpressionLCFT}
        G_{12} = \frac{1}{2\nu} \frac{d}{d \nu} G_{22}
        = \frac{1}{2\nu} \frac{d}{d \nu}\left( \frac{u^{1+d/2} K_{\nu} (q u)}
        {u_{B}^{1+d/2} K_{\nu} (q u_{B})} \right) \;.
    \end{align}

After plugging the bulk equation of motion into the action (\ref{SchrActionLCFT}),
the boundary action becomes of the form
	\begin{align}   \label{boundaryActionLCFT}
		S_B &= -K \int  d^{d+1} x \frac{L^{d+3}}{u^{d+3}} ~\tilde \phi  \frac{u^2}{L^2} \partial_u
			 \phi \Big|_{u_B} \nonumber \\
			&= -K \int [t, \omega', \omega] \int [x, \vec k', \vec k]
			~~\tilde J (k') \Big[ \mathcal F_1 (u_B, k', k ) J (k)
			+ \mathcal F_2 (u_B, k', k) \tilde J (k) \Big] \;.
	\end{align}
For ALCF, the system has the time translation invariance, thus the time integral is
trivially evaluated to give a delta function.
The $\mathcal F$'s are given by
	\begin{align}    \label{F1Function2} 
		\mathcal F_1	&= \frac{L^{d+1}}{u^{d+1}}G_{22} (k')~ \partial_u ~ G_{11} (k)  \;, \\
		\mathcal F_2	&= \frac{L^{d+1}}{u^{d+1}}G_{22} (k') ~\partial_u ~ G_{12} (k)   \;.
		\label{F2Function}
	\end{align}
We note that $\mathcal F_1$ leads the same result as (\ref{F1Function}) and (\ref{F1Correlator}). 

Let us evaluate $\mathcal F_2$, which is 
$\nu$ derivative of $\mathcal F_1$ given in (\ref{G12ExpressionLCFT}). The result is  
	\begin{align}  \label{F2Correlator}
		\langle \phi (x_{2}) \phi (x_{1}) \rangle_{\mathcal F_2}
		&= \frac{1}{2 \Gamma (\nu+1)} 
		\frac{L^{d+1} \mathcal M^{\frac{d}{2}+\nu}  }{\pi 2^{\nu-1}  u_B^{d+2-2\nu}} 
\cdot \frac{\theta (t_2) \theta(t_2 - t_1) }{ (t_2 - t_1)^{\frac{d+2}{2}+\nu}}   
		\cdot \exp \left( -{ \frac{\mathcal M (\vec{x}_2 - \vec{x}_1)^2 }{2 ( t_2 - t_1)}}\right) \nonumber \\
		&\quad \times \left( 1 -\nu \frac{\Gamma (\nu)'}{\Gamma (\nu)} 
		+ \nu \ln [\frac{\mathcal M u_B^2}{2(t_2 - t_1)}]
		 \right) \;.
	\end{align}
These are our correlation and response functions with log extensions for 
general $z$ and $d$. This is a direct 
generalization of the previous result for $d=2$ and $z=2$ \cite{Hyun:2012fd}. 
Note that the result is valid for the field theories with Galilean boost without 
conformal symmetry similar to the result (\ref{F1Correlator}). 
Again it is independent of $z$.

\subsubsection{Response functions with log$^2$ extensions}   \label{sec:ALCFCorrLog2}

Motivated by the recent interests on tricritical log gravity \cite{Bergshoeff:2012ev}, 
we consider three scalar fields $\phi_1$, $\phi_2$ and $\phi_3$ in the background 
(\ref{ALCFMetric}) with the following action 
	\begin{align}  \label{SchrActionTLCFT}
		S = K \int d^{d+2} x \int_{u_B}^{\infty} du \sqrt{-g}
		& \left[ \frac{1}{2} \partial^M \phi_2 \partial_M \phi_2 
		+ \partial^M \phi_3 \partial_M \phi_1
		+ m^2 \left[ \frac{\phi_2^2}{2} + \phi_3 \phi_1 \right] +\frac{\phi_3 \phi_2}{2L^2}  \right]  \;,
	\end{align}
where we take $\partial_\xi \phi_i =-\mathcal M \phi_i$ for $i=1,2,3$.  
This action is previously considered in \cite{Moon:2012vc} in a different context. 
The field equations for $\phi$'s of the action (\ref{SchrActionLCFT}) become
	\begin{align}   \label{DiffEqSchrTLCFT}
		&2{\cal M} \partial_t \phi_1 =
		- \frac{1}{u^2} \mathcal D  \phi_1 - \vec \nabla ^2  \phi_1 + \frac{1}{u^2} \phi_2 \;, \\
		&2{\cal M} \partial_t \phi_2 =
		- \frac{1}{u^2} \mathcal D  \phi_2 - \vec \nabla ^2  \phi_2 + \frac{1}{u^2} \phi_3 \;, \\
		&2{\cal M} \partial_t \phi_3 =
		- \frac{1}{u^2} \mathcal D \phi_3 - \vec \nabla ^2 \phi_3  \;,
	\end{align}
where $\mathcal D$ is given in (\ref{DiffOpALCF}).

We construct the bulk to boundary Green's function $G_{ij} (u, \omega, \vec{k})$ in 
terms of $J_i (  \omega, \vec{k})$ as
	\begin{align}   \label{BulkToBoundaryGreensFunctionTLCFT}
		&\phi_i (u,x^\mu) = \int \frac{d^d k}{(2\pi )^d}\frac{d \omega}{2 \pi} e^{i k \cdot x }
		 G_{ij} (u,k) J_j (k) \;, 
	\end{align}
We choose $G_{21}= G_{31}=G_{32}=0$, which is in accord with the structure of the equations of motion given
in (\ref{DiffEqSchrTLCFT}). The Green's functions satisfy
\begin{align}
\begin{array}{ccc}
	\mathrm D G_{11} = 0  \;, \qquad
	&\mathrm D G_{12} = G_{22}   \;, \qquad
	&\mathrm D G_{13} = G_{23}   \;, \\
	\mathrm D G_{21} = 0  \;, \qquad
	&\mathrm D G_{22} = 0   \;, \qquad
	&\mathrm D G_{23} = G_{33}  \;, \\
	\mathrm D G_{31} = 0  \;, \qquad
	&\mathrm D G_{32} = 0  \;, \qquad
	&\mathrm D G_{33} = 0   \;, 
\end{array} 
\end{align}
where $\mathrm D = \mathcal D - q^2 u^2 $ and $q = \sqrt{\vec k^2 +2 {\cal M} i\omega}$.
The Green's functions $G_{ii}, i=1, 2, 3$ are 
	\begin{align}		
		&G_{ii}(u, k)=  c_{ii}~u^{1+d/2} K_\nu (q u)\;, 
	\end{align}
where $\nu = \pm \sqrt{(1+d/2)^2 + m^2L^2 } $ with the same normalization constant 
given in (\ref{OnShAct1}).

There exist other Green's functions $G_{12}, G_{13}, G_{23}$ for the action (\ref{SchrActionTLCFT}),
which satisfies
	\begin{align}
	\mathrm D G_{12} = G_{22} \;, \qquad \mathrm D G_{23} = G_{33} \;, \qquad 
	\mathrm D G_{13} = G_{23}   \;.
	\end{align}
In particular, we have 
	\begin{align}
	\mathrm D^2 G_{13} = G_{33}  \;, \quad 
	\mathrm D^3 G_{13} = \mathrm D^2 G_{23} = \mathrm D G_{33} =0   \;.
	\end{align}
To evaluate them, we generalize the methods used in \cite{Kogan:1999bn}\cite{Hyun:2012fd} 
to the tricritical case.
Using again $\left[ \mathrm D , d / d \nu \right]  = 2 \nu $,
and the fact that $ \mathrm D G_{ii}=0 $, we get
    \begin{align}
        \mathrm D \left( \frac{1}{2 \nu} \frac{d}{d\nu} G_{ii} \right) =G_{ii} \;.
    \end{align}
Thus
	\begin{align}   
        &G_{12} = \frac{1}{2\nu} \frac{d}{d \nu} G_{22}
        = \frac{1}{2\nu} \frac{d}{d \nu}\left( \frac{u^{2} K_{\nu} (q u)}
        {u_{B}^{2} K_{\nu} (q u_{B})} \right) \;,  \\
        &G_{23} = \frac{1}{2\nu} \frac{d}{d \nu} G_{33}
        = \frac{1}{2\nu} \frac{d}{d \nu}\left( \frac{u^{2} K_{\nu} (q u)}
        {u_{B}^{2} K_{\nu} (q u_{B})} \right) \;, \\
        &G_{13} = \frac{1}{4\nu} \frac{d}{d \nu} G_{23}
        = \frac{1}{4\nu} \frac{d}{d \nu}\left( \frac{1}{2\nu} \frac{d}{d \nu} G_{33} \right) \;. 
        \label{F3Function}
    \end{align}
Note that the last expression has second order derivative of $\nu$, which leads log$^2$ contributions. 

After plugging the bulk equation of motion into the action (\ref{SchrActionTLCFT}),
the boundary action becomes of the form
	\begin{align}   \label{boundaryActionTLCFT}
		S_B &= -K \int  d^{d+1} x \frac{L^5}{u^{5}} ~\left( \phi_2  \frac{u^2}{L^2} \partial_u \phi_2 
		+  \phi_3  \frac{u^2}{L^2} \partial_u \phi_1  \right) \bigg|_{u_B} \nonumber \\
			&= -K \int [t,\omega',\omega]  \int[\vec x, \vec k', \vec k]
			~~ J_i (k'_\mu) \mathcal F_{ij} (u_B, k'_\mu, k_\mu ) J_j (k_\mu)  \;.
	\end{align}
The system has time translation invariance, thus the time integral is
trivially evaluated to give delta function.
The $\mathcal F$'s are given by
	\begin{align}    \label{FsTLCFT}
		\mathcal F_{22}	&= \frac{L^3}{u^3}G_{22} (k'_\mu)~ \partial_u ~ G_{22} (k_\mu) = \mathcal F_1 \;, \\
		\mathcal F_{23}	&= \frac{L^3}{u^3}G_{22} (k'_\mu) ~\partial_u ~ G_{23} (k_\mu) = \mathcal F_2  \;, \\
		\mathcal F_{31}	&= \frac{L^3}{u^3}G_{33} (k'_\mu)~ \partial_u ~ G_{11} (k_\mu) = \mathcal F_1 \;, \\
		\mathcal F_{32}	&= \frac{L^3}{u^3} \left( G_{23} (k'_\mu) ~\partial_u ~ G_{22} (k_\mu)  
		+ G_{33} (k'_\mu) ~\partial_u ~ G_{12} (k_\mu) \right) =  \mathcal F_2\;, \\
		\mathcal F_{33}	&= \frac{L^3}{u^3} \left( G_{23} (k'_\mu)~ \partial_u ~ G_{23} (k_\mu) 
		+ G_{33} (k'_\mu)~ \partial_u ~ G_{13} (k_\mu)  \right) = \mathcal F_3	\;. \label{LastFsTLCFT}
	\end{align}
Note that the first terms in $ \mathcal F_{32}$ and  $\mathcal F_{33} $ are $0$ when evaluated at $u=u_B$. 
We also notice that $ \mathcal F_1 $ and $ \mathcal F_2 $ are identical to (\ref{F1Function}) and
(\ref{F2Function}), and thus the corresponding correlation functions 
(\ref{F1Correlator}) and (\ref{F2Correlator}), respectively. 

Now we are ready to evaluate $\mathcal F_3 $. Using (\ref{F3Function}) and 
(\ref{F1Correlator}), we get 
	\begin{align}  \label{F3Correlator}
		\langle \phi_3 (x_{2}) \phi_3 (x_{1}) \rangle_{\mathcal F_3}
		&= \frac{1}{8 \nu^3 \Gamma (\nu)} 
		\frac{L^{d+1} \mathcal M^{\frac{d}{2}+\nu}  }{\pi 2^{\nu-1}  u_B^{d+2-2\nu}} 
		\cdot \frac{\theta (t_2) \theta(t_2 - t_1) }{ (t_2 - t_1)^{\frac{d+2}{2}+\nu}}  
		\cdot \exp \left( -{ \frac{\mathcal M (\vec{x}_2 - \vec{x}_1)^2 }{2 ( t_2 - t_1)}}\right)  \\
		& \times  \left[ 1 +\nu \psi - \nu^2( \psi^2- \psi') 
		+( 2\nu^2 \psi - \nu) \ln [\frac{\mathcal M u_B^2}{2(t_2 - t_1)}] 
		- \nu^2 \ln [\frac{\mathcal M u_B^2}{2(t_2 - t_1)}]^2  \right]  \;,  \nonumber
	\end{align}
where $\psi(\nu) = \frac{\Gamma (\nu)'}{\Gamma (\nu)} $.
This is our main result in this section, response functions with log and log$^2$ extensions, 
which is valid for general $z$ and $d$. 
Note also that this result is valid for the systems without non-relativistic conformal invariance. 
We notice that various coefficients in the square bracket are completely determined once $\nu$ is fixed.

\subsection{Schr\"odinger backgrounds}  \label{sec:Schr}

We first establish the Schr\"odinger type solutions with Galilean symmetry 
with $z\neq 2$ following \cite{Balasubramanian:2008dm}, 
see also \cite{Kim:2012nb}\cite{Son:2008ye}. 
Finite temperature generalizations for general $z$ is considered in \cite{Kim:2012pd}, 
while those for $z=2$ are considered in 
\cite{Herzog:2008wg}\cite{Adams:2008wt}\cite{Yamada:2008if}\cite{Mazzucato:2008tr}\cite{Ammon:2010eq}. 

The metric at zero temperature is given by 
	\begin{align}   \label{SchrMetric}
		ds^2 = L^2 \left(-\gamma \frac{dt^2}{u^{2z}} +\frac{d\vec x^2 - 2 dt d\xi + du^2}{u^{2}} \right) \;,  
	\end{align}
which is invariant under the space-time translations $P_i, H$, Galilean boost $K_i$, scale transformation $D$ 
and translation along the $ \xi $ coordinate. Their explicit forms are given in (\ref{GalileanTR1}) and (\ref{ScaleTR1}). 
There exists additional special conformal transformation for $z=2$, which has been focus of 
previous investigations.

There have been more general class of gravity backgrounds with so-called hyperscaling violation. 
These backgrounds are described by $ds^2 = u^{2-2\frac{\theta}{d+1}} 	\left[-\gamma \frac{dt^2}{u^{2z}} 
+\frac{d\vec x^2 - 2 dt d\xi + du^2}{u^{2}} \right]  $ considered in \cite{Kim:2012nb}, 
where $\theta$ is a hyperscaling violation exponent. 
$\theta$ is first introduced in \cite{Gouteraux:2011ce} based on \cite{CGKKM}. 
This hyperscaling violation might be also interesting in the general context of aging and growth phenomena. 
The associated matter fields are a gauge field, a scalar and the non-trivial coupling between them. 

The geometry (\ref{SchrMetric}) is not a solution of vacuum Einstein equations. 
Thus we require to support it with some matter fields. One particular example is the ground state 
of an Abelian Higgs model in its broken phase \cite{Balasubramanian:2008dm}
	\begin{align}
		&S = \int d^{d+3}x \sqrt{-g} \left(-\frac{1}{4} F^2 - \frac{1}{2} |D \Phi |^2 - V(|\Phi |) \right) \;, \\
		&V(|\Phi |) =  \left( |\Phi |^2 -v^2 \right)^2 -\frac{(d+1)(d+2)}{L^2} \;,  \\
		&A_t (u) = \frac{\rho_0}{ u^{z}}  \;, \quad 
		\rho_0^2 = \frac{2 L^2  (z-1)(2z+d) \gamma }{z^2 + e^2 v^2 L^2 } \;, 
	\end{align}
where $F =dA$, $F^2 = F^{MN} F_{MN} $, and $d$ is the number of spatial dimensions.

It is not hard to find a different matter system that supports the metric \cite{Son:2008ye}.
	\begin{align}
		&S = \int d^{d+3}x \sqrt{-g} \left(-\frac{1}{4} F^2 -\frac{m^2}{2} A^2 + \frac{(d+1)(d+2)}{L^2} \right) \;, \\
		&A_t (u) = \frac{\rho_0}{ u^{z}}  \;, \quad \rho_0^2 
		=  \frac{2 \left(d+2z\right)(z-1)L^2 \gamma }{3z} \;,  \quad m^2 = \frac{z(3-z)}{L^2} \;,
	\end{align}
where $ A^2 = A^M A_M$.

\subsubsection{Correlation and response functions with log $\&$ log$^2$ extensions}   
\label{sec:SchrLog}

We are interested in constructing correlation and response functions using three different 
actions, (\ref{ScalarActionALCF}), (\ref{SchrActionLCFT}) and (\ref{SchrActionTLCFT}), 
as in the previous section \S \ref{sec:ALCF}. Here we briefly show that the procedure 
is the same as before. Thus we can compute the logarithmic (squared) extensions by taking 
a simple $\nu$ derivatives of the correlation function obtained from the action 
(\ref{ScalarActionALCF}). 

We start by considering correlation functions of the geometry (\ref{SchrMetric}) 
by coupling a probe scalar with the same action as (\ref{ScalarActionALCF}). 
The field equation for $\phi$ becomes
\begin{align}
 \frac{\partial^2 \phi}{\partial u^2}  -(d+1) \frac{1}{u} \frac{\partial \phi}{\partial u} 
-\left ( \frac{ m^2 L^2 }{u^2} +\vec k^2 +2 M w +\gamma M^2 u^{2-2z} \right) \phi = 0
  \;.
\label{BlukScalarEqSchr}
\end{align} 
Note $\gamma \neq 0$, which is one of the main differences between the Schr\"odinger background 
(\ref{SchrMetric}) and ALCF (\ref{ALCFMetric}). 
Again, we treat $\xi$ coordinate special and replace all the $\partial_{\xi}$ as $iM$. 
With this differential equation, one can compute the correlation function $\mathcal F_1$. 
For general $z$, analytic solutions are not available.
The resulting correlation function for $z=2$ is already computed in 
\cite{Jottar:2010vp}\cite{Hyun:2011qj}\cite{Hyun:2012fd}, while that of $z=3/2$ is 
computed below in \S \ref{sec:zThreeHalf}. Previously, several special cases also have been 
computed in \cite{Kim:2012nb}. 

To compute the corresponding log and log$^2$ extensions, we consider a
Schr\"odinger differential operator 
\begin{align}   \label{DiffOpSchr}
	\mathrm D_{Schr} = u^2 \bigg[ \partial_u^2 - \frac{d+1}{u} \partial_u
		- \frac{\nu^2 -(1 +d/2)^2 }{u^2} - q^2 -\gamma M^2 u^{2-2z} \bigg] \;, 
\end{align} 
where $ q^2 = \vec k^2 + 2M\omega$ and $\nu = \pm \sqrt{(1+d/2)^2+ L^2 m^2}$ for $z\neq 2$. 
For the special case $z=2$, we have $\nu = \pm \sqrt{(1+d/2)^2+ L^2 m^2 + \gamma M^2}$ from 
 (\ref{DiffOpSchr}). 
With the differential operator $ \mathrm D_{Schr} $, we can still use the relation 
    \begin{align}
        \left[ \mathrm  D_{Schr} \;, \frac{d}{d \nu} \right]  = 2 \nu \;,
    \end{align}
to compute the correlation (response) functions with the logarithmic extensions. 
For that purpose, we use the equations (\ref{BulkToBoundaryGreensFunctionLCFT}) 
- (\ref{F1Function2}) with appropriate $G_{11} $ and $G_{22} $. 
We also get the response functions with the log$^2$ extension using 
(\ref{BulkToBoundaryGreensFunctionTLCFT}) 
- (\ref{LastFsTLCFT}) with appropriate $G_{11}, G_{22} $ and $G_{33} $. 

The upshot is that the logarithmic extensions can be computed by taking one or two derivatives 
of the correlation functions available. 
Let us compute these correlation and response functions for $z=2$ and $z=3/2$ in turn.

\subsubsection{Conformal Schr\"odinger backgrounds with $z=2$}  \label{sec:zTwo}

We comment for the conformal case $z=2$ here. 
The differential equation (\ref{BlukScalarEqSchr}) simplifies to 
\begin{align}
 \frac{\partial^2 \phi}{\partial u^2}  -(d+1) \frac{1}{u} \frac{\partial \phi}{\partial u} 
-\left ( \frac{ m^2 L^2 + \gamma M^2 }{u^2} +\vec k^2 +2 M w \right) \phi = 0   \;.
\end{align} 
This is similar to that of ALCF given in (\ref{BlukScalarEqALCF}), 
the only difference is the presence of the parameter $\gamma$, which modifies $\nu$ as  
$\nu = \pm \sqrt{(1+d/2)^2 + m^2 L^2 - \gamma \mathcal M^2}$. 
This observation leads us that we can compute correlation and response functions with 
logarithmic extensions as in \S \ref{sec:ALCF}. These are 
(\ref{F1Correlator}),  (\ref{F2Correlator}) and (\ref{F3Correlator}) with modified $\nu$.

\subsubsection{Schr\"odinger backgrounds with $z=3/2$}  \label{sec:zThreeHalf}

The case $z=\frac{3}{2}, d=1$ is our main interest for the application to KPZ universality class. 
For the time being, we work on general spatial dimensions $d$. The corresponding solution is 
\begin{align}  \label{Schr32Sol}
	\phi = e^{-qu} u^{\frac{2+d}{2}+\nu}\left[ 
	c_1  U \left( a,1+2\nu,2 qu \right)
	+c_2  L_{-a}^{2\nu}( 2 q u ) \right] \;, 
\end{align}
where $a= \frac{M^2}{2q} +\frac{1+2\nu}{2}$, $\gamma=1$, $\nu = \pm \sqrt{(1+d/2)^2+ L^2 m^2}$ and 
$ q^2 = \vec k^2 + 2M\omega$. $U$ and $ L$ represent the
confluent hypergeometric function and the generalized Laguerre polynomial. 
We choose $U$ for our regular solution. 

The momentum space correlation function can be evaluated as the ratio between the 
normalizable and non-normalizable contributions at the boundary expansion of the solution 
(\ref{Schr32Sol}), which is given by \cite{Kim:2012nb}
\begin{align}   \label{SchrMomentumCorr}
	G(q) \sim  4^{\nu } q^{2 \nu } \frac{\Gamma [-2 \nu ] }{\Gamma [2 \nu ]}
	 \frac{\Gamma \left[\frac{1 + 2\nu}{2}+\frac{M^2}{2 q}\right]}
	 {\Gamma \left[\frac{1 - 2\nu}{2}+\frac{M^2}{2 q}\right]} \;, 
\end{align}
where we only keep momentum dependent parts. One can restore the $u_B$ dependence 
using scaling arguments. 
For the general case, Fourier transforming back analytically to the coordinate space is difficult. 
Thus we would like to consider some special cases. 

\begin{itemize}
\item[A.] $\frac{M^2}{2 q} \rightarrow 0 $ : 
The momentum space correlator has the same form as aging in ALCF
\begin{align}
	G(q) \sim 4^{\nu } q^{2 \nu }  \frac{\Gamma [-2 \nu ] }{\Gamma [2 \nu ]}
	 \frac{\Gamma \left[\frac{1 + 2\nu}{2}\right]}{\Gamma \left[\frac{1 - 2\nu}{2}\right]} \;. 
\end{align}
For imaginary parameter $M = i \mathcal M$, we get 
\begin{align}
&\langle \phi^* (x_{2})  \phi(x_{1}) \rangle _{M_ 0} 
\sim \frac{1}{\Gamma (\nu)} \frac{\mathcal M^{\frac{d}{2} + \nu}}{\pi 2^{\nu+1}}
\frac{\theta (t_2) \theta(t_2 - t_1) }{ (t_2 - t_1)^{\frac{d+2}{2}+\nu}}   
\cdot \exp \left( -{ \frac{\mathcal M (\vec{x}_2 - \vec{x}_1)^2 }{2 ( t_2 - t_1)}}\right)  \;. 
\label{SchrCorrF1}
\end{align}
This result is for $z=\frac{3}{2}$. The dependence on time and space is identical to the 
result of the aging in ALCF. 

We are interested in log and log$^2$ extensions. For this purpose, we consider 
\begin{align}
&\langle \phi^* (x_{2}) \phi (x_{1}) \rangle _{\mathcal F_1}^{M_ 0} 
\sim h(\nu) \left( \frac{\mathcal M u_B^2}{t_2 - t_1} \right)^\nu   \;, 
\end{align}
where $h(\nu)$ collectively denotes the other $\nu$ dependent parts. 
We also use the same actions (\ref{SchrActionLCFT}) and (\ref{SchrActionTLCFT}) to get the 
correlation functions with the logarithmic extension 
	\begin{align}   \label{SchrCorrF2}
		&\langle \phi (x_{2}) \phi (x_{1}) \rangle_{\mathcal F_2}^{M_ 0} 
		= \frac{1}{2\nu} \left( \frac{h (\nu)'}{h (\nu)} 
		+  \ln [\frac{\mathcal M u_B^2}{(t_2 - t_1)}]
		 \right) 
		 \langle \phi (x_{2}) \phi (x_{1}) \rangle _{\mathcal F_1}^{M_0}  \;,
	\end{align}
and the log$^2$ extension 
	\begin{align}  \label{SchrCorrF3}
		&\langle \phi_3 (x_{2}) \phi_3 (x_{1}) \rangle_{\mathcal F_3}^{M_0}  
		=\frac{1}{8\nu^3} 
		 \left[ A_0  +A_1 \ln [\frac{\mathcal M u_B^2}{(t_2 - t_1)}] 
		+ A_2 \ln [\frac{\mathcal M u_B^2}{(t_2 - t_1)}]^2  \right] 
		 \langle \phi _2 \phi_1 \rangle _{\mathcal F_1}^{M_ 0}  \;,
	\end{align}
where $A_0 =\frac{-h'[\nu ]+\nu  h''[\nu ]}{h[\nu ]} $, 
$ A_1 =\frac{-h[\nu ]+2 \nu  h'[\nu ]}{h[\nu ]}$ and $ A_2  =\nu $.

\item[B.] $\frac{M^2}{2 q} \rightarrow \infty$ : 
We use the asymptotic expansion form from \S 5.11 of \cite{NIST}
\begin{align}
\lim_{z \rightarrow \infty} \frac{\Gamma [z+a] }{\Gamma [z+b]} \sim z^{a-b} \sum_{k=0}^{\infty} 
\frac{G_k (a,b)}{z^k} \;, \quad 
G_{k}(a,b)=\binom{a-b}{k} B^{{(a-b+1)}}_{{k}} (a) \;,
\end{align} 
where $ \binom{a-b}{k} $ are binomial coefficients and 
$B^{{(a-b+1)}}_{{k}} (a) $'s are generalized Bernoulli polynomials. 
For our case, $ G_{2n-1} \propto (a+b-1) = 0 $. 

The momentum space correlation function is 
\begin{align}
	G(q) &\sim 4^{\nu } q^{2 \nu } \frac{\Gamma [-2 \nu ] }{\Gamma [2 \nu ]}
	\left( \frac{M^2}{2 q} \right)^{2\nu} \sum_{n=0}^{\infty} G_{2n} \left(\frac{q}{M^2} \right)^{2n} \;,
\end{align}
The first term is independent of momenta, which we ignore. 
For the rest of the terms, the inverse Fourier transform of $q^{2n}$ gives us $\frac{1}{\Gamma(-n)}$, 
which vanishes for integer $n$. 
Thus the coordinate correlation function identically vanishes except the case $n=2\nu $.
Thus we get, using $ M= i \mathcal M $ 
\begin{align}     \label{SchrCorrLargeMF1}
&\langle \phi^* (x_{2})  \phi (x_{1}) \rangle _{M_\infty} 
\sim  \frac{G_{4\nu}}{\Gamma [2 \nu ] } 
\frac{\mathcal M^{\frac{d}{2}-2\nu}}{\pi 2^{1-2\nu} }
\cdot \frac{\theta (x_2^+) \theta(t_2 - t_1) }{ (t_2 - t_1)^{\frac{d+2}{2}+2\nu}}   
 \cdot \exp \left( -{ \frac{\mathcal M (\vec{x}_2 - \vec{x}_1)^2 }{2 ( t_2 - t_1)}}\right)  \;. 
\end{align}

We are interested in the response functions with log and log$^2$ extensions, we consider 
\begin{align}
&\langle \phi^* (x_{2})  \phi (x_{1}) \rangle _{\mathcal F_1}
^{M_\infty}
\sim \tilde h(\nu) \left( \frac{\mathcal M u_B^2}{t_2 - t_1} \right)^{2\nu}   \;. 
\end{align}
where $\tilde h(\nu)$ collectively denotes the other $\nu$ dependent parts. 
We also use the same actions (\ref{SchrActionLCFT}) and (\ref{SchrActionTLCFT}) to get the 
correlation functions with the logarithmic extension 
	\begin{align}   \label{SchrCorrLargeMF2}
		&\langle \phi (x_{2}) \phi (x_{1}) \rangle_{\mathcal F_2}^{M_ \infty}
		= \frac{1}{2\nu} \left( \frac{\tilde h (\nu)'}{\tilde h (\nu)} 
		+ 2 \ln [\frac{\mathcal M u_B^2}{(t_2 - t_1)}] \right) 
		 \langle \phi^* \phi \rangle _{\mathcal F_1}^{M_\infty} \;,
	\end{align}
and the log$^2$ extension 
	\begin{align}  \label{SchrCorrLargeMF3}
		&\langle \phi_3 (x_{2}) \phi_3 (x_{1}) \rangle_{\mathcal F_3}^{M_\infty} 
		=\frac{1}{8\nu^3} 
		 \left[ \tilde A_0	+\tilde A_1 \ln [\frac{\mathcal M u_B^2}{(t_2 - t_1)}] 
		+ \tilde A_2 \ln [\frac{\mathcal M u_B^2}{(t_2 - t_1)}]^2  \right] 
		 \langle \phi^*  \phi \rangle _{\mathcal F_1}^{M_\infty}  \;,
	\end{align}
where $\tilde A_0=\frac{-\tilde h'[\nu ]+\nu  \tilde h''[\nu ]}{\tilde h[\nu ]} $, 
$\tilde  A_1 =\frac{-2\tilde h[\nu ]+4 \nu  \tilde h'[\nu ]}{\tilde h[\nu ]}$ 
and $\tilde A_3 =4 \nu $. 

\end{itemize}
These two extreme cases, $\mathcal M \rightarrow 0$ and $\mathcal M \rightarrow \infty $, 
signal that the parameter $\mathcal M$ can bring some quantitatively different behaviors of 
the correlation and response functions because of the different power in time dependent denominators 
$ (t_2 - t_1)^{-\frac{d+2}{2}-\nu}$ and $ (t_2 - t_1)^{-\frac{d+2}{2}-2\nu}$ in 
(\ref{SchrCorrF1}) and (\ref{SchrCorrLargeMF1}), respectively.

\subsection{Two-time response functions}    \label{sec:NTLFTsummary}

In this section we summarize \S \ref{sec:NTLFT} by considering the two-time correlation and 
response functions with logarithmic extensions. From the various results of ALCF and 
Schr\"odinger backgrounds, equations (\ref{F3Correlator}), (\ref{SchrCorrF3}) 
and (\ref{SchrCorrLargeMF3}), we observe that the correlation functions with(out) log extensions 
show qualitatively similar properties. 

Some typical two-time correlation and response functions can be obtained by putting 
$ \vec x_2 = \vec x_1$ in equation (\ref{F3Correlator}). 
	\begin{align}  
		& C(t_2 , t_1) =	
		(t_2 - t_1)^{-\frac{d+2}{2}-\nu} \left[ A_0 
		+A_1 \ln [\frac{\mathcal M_B}{t_2 - t_1}] 
		+ A_2 \ln [\frac{\mathcal M_B}{t_2 - t_1}]^2  \right]  \;,
	\end{align}
where $\psi(\nu) = \frac{\Gamma (\nu)'}{\Gamma (\nu)}, 
\mathcal M_B = \frac{\mathcal M u_B^2}{2}$, and the coefficients 
\begin{align}
&A_0 =1 +\nu \psi - \nu^2( \psi^2- \psi')  \;, \\
&A_1 =( 2\nu^2 \psi - \nu) \;, \qquad  A_2 =- \nu^2 \;.
\end{align}
We note that these coefficients, $A_0, A_1 $ and $A_2$, are determined once $\nu$ is fixed. 
$C(t_2, t_1)$ is invariant under the time translation transformation, 
and so $C(t_2, t_1)=C(t_2- t_1)$. The so-called ``waiting time'' 
$s=t_1$ does not have a physical meaning. Thus $C(t_2 - t_1)$ is completely fixed as a function of 
$t_2 - t_1$, once $d$, $\nu$ and $\mathcal M u_B^2$ are given. 
Physically, this time translation invariant two-time response functions describe 
either constant growth or constant aging (decaying) phenomena.
Further physical significances are considered in detail in \S \ref{sec:KPZ}.

\section{Aging Logarithmic Galilean Field Theories}  \label{sec:Aging}

Equipped with the generalization of our correlation and response functions for general $z$ and $d$, 
non-relativistic and (non-)conformal geometries, we would like to add yet another ingredient 
to them : aging, one of the simplest time-dependent physical phenomena.  
Typically aging is realized when the system is rapidly brought out of equilibrium.
For this simple time-dependent phenomena, time translational invariance is broken. 
There are two important time scales: (1) {\it waiting time} which marks the time scale when 
the system is perturbed after it is put out of equilibrium and (2) {\it response time} 
which marks when the perturbation is measured. 
Typical properties of aging are described by the two-time response functions in terms of 
these waiting time and response time, and are power law decay, broken time translation 
invariance and dynamical scaling between the time and spatial coordinates. 
These are shown in holographic model in \cite{Hyun:2011qj} as well as 
various field theoretical models, see {\it e.g.}  
\cite{HenkelBook2}\cite{HinrichsenReviewDP}\cite{Henkel:2007nept}.

In the context of Anti-de Sitter space/Conformal field theory correspondence (AdS/CFT) 
\cite{Maldacena:1997re}\cite{Aharony:1999ti} and its extension to Schr\"odinger geometries 
\cite{Son:2008ye}\cite{Balasubramanian:2008dm}\cite{Goldberger:2008vg}\cite{Barbon:2008bg},
the geometric realizations of aging have been put forward in 
\cite{Jottar:2010vp}\cite{Hyun:2011qj} 
by generalizing the background with explicit time dependent terms.
These terms are generated by a singular time dependent coordinate transformation, 
which itself has significant physical meaning in the context of holography \cite{Jottar:2010vp}. 
Furthermore, there exists a time boundary at $t=0$ and physical boundary conditions are 
explicitly imposed: (1) by complexifying time in \cite{Jottar:2010vp} or 
(2) by introducing some decay modes of the bulk scalar field along the `internal' 
spectator direction $\xi$, which is not explicitly visible from the dual
field theory in \cite{Hyun:2011qj}. 
We prefer the option (2) in this paper as in \cite{Hyun:2011qj}, where 
the resulting two-time correlation functions show a dissipative
behavior and exhibit the three characteristic features of the aging system mentioned above. 
Thus the time translation symmetry is broken globally, and the aging
symmetry is realized as conformal Schr\"odinger symmetry modulo time translation symmetry 
\cite{Jottar:2010vp}\cite{Hyun:2011qj}. 
Their finite temperature properties with asymptotic aging invariance are also 
investigated in \cite{Hyun:2011qj}. See also a recent review \cite{Gray:2013rv}. 

In this section we would like to generalize this aging construction to the case with general 
dynamical exponent $z$ and for general dimensions $d$. 
The generalization of the singular coordinate transformation and 
the corresponding aging geometries are constructed in \S \ref{sec:AgingTransform}. 
In \S \ref{sec:agingALCF}, we construct the two point correlation and response functions 
for ALCF in the context of \cite{Goldberger:2008vg}\cite{Barbon:2008bg}, while similarly 
in \S \ref{sec:agingBack} for Schr\"odinger background in the context of 
\cite{Son:2008ye}\cite{Balasubramanian:2008dm}. 
Their log and log$^2$ extensions are explained in \S \ref{sec:AgingLogExtension}.

\subsection{Constructing aging geometry for general $z$}  \label{sec:AgingTransform}

Physical properties of aging is explored in holography by using a {\it singular} coordinate 
transformation 
	\begin{equation}
		\xi \quad \longrightarrow \quad  \xi - \frac{\alpha}{2}  \ln \left( u^{-2} t \right) \;,
		\label{CoordinateChangezTwo}
	\end{equation}
which is first introduced in \cite{Jottar:2010vp}, specifically for $z=2$ case. 
It is important to impose physical boundary conditions on the time boundaries in addition to the 
spatial boundaries. The simplest possibility in this context has been explored in \cite{Hyun:2011qj}

We would like to extend this singular transformation for general $z$ in a direct manner. 
	\begin{equation}   \label{CoordinateChangez}
		\xi \quad \longrightarrow \quad  \xi - \frac{\alpha}{2}  \ln \left( u^{-z} t \right) \;.
	\end{equation}
Note that for general $z$, the coordinate $\xi$ has non-trivial dimensions, $[\xi] = 2-z $, 
under the scaling transformation. One immediate consequence is a nontrivial scaling dimension of 
our parameter $[\mathcal M] = z-2$. This is already observed in the exponent of the 
correlation and response functions in previous section. 
Now for the aging extension, we observe that the parameter $\alpha$ also has a definite 
scaling dimension $[\alpha] = 2-z$. These two parameters conspire to provide us a rather simple 
and elegant generalization to the aging correlation and response functions for general $z$. 

The background metric extended to the aging is correspondingly modified to  
	\begin{align}   \label{AgingMetric}
		ds_u^2 =   \frac{L^2}{u^2} \left(  d\vec{x}^2  -2 dt d\xi
		- \left(\frac{\gamma}{u^{2z-2}} +  \frac{\alpha}{t} \right) dt^2
			+ \frac{z \alpha  }{u} du dt  +  du^2  \right) \;,
	\end{align}
where $\gamma = 0$ corresponds to aging in ALCF. There exists a slight change in metric 
compared to (\ref{ALCFMetric}) or (\ref{SchrMetric}) :   
the coefficient of the term $ du dt  $ has a factor of $z$ instead of $2$. 
One can check that the matter contents without the singular transformation would solve the 
corresponding Einstein equation. These cases can be considered as {\it locally Galilean}.  

To compute the correlation functions of the probe scalar fields in the background geometry 
(\ref{AgingMetric}) with general $z$ and $d$, we consider the action given in (\ref{ScalarActionALCF}).    
The field equation for $\phi$ becomes
\begin{align} \label{BlukScalarEqAgingALCF}
&2M \left[ i \frac{\partial}{\partial t} + \frac{\alpha M}{2 t} \right] \phi  \\
& = \frac{\partial^2 \phi}{\partial u^2}  + \frac{z i M\alpha -d-1}{u} \frac{\partial \phi}{\partial u} 
-\left[ \frac{ 4 m^2 L^2 + 2(d+2) z i M \alpha + z^2 M^2 \alpha^2  }{4 u^2}  
+ \frac{\gamma M^2}{ u^{2z-2}} +\vec k^2  \right] \phi \;. \nonumber 
\end{align}
Note that here we treat $\xi$ coordinate special and replace all $\partial_{\xi}$ as $iM$, 
because this coordinate plays a distinguished role in Galilean and corresponding aging holography 
\cite{Balasubramanian:2008dm}\cite{Kim:2012nb}. 

To find the solution of the equation (\ref{BlukScalarEqAgingALCF}), 
we use the Fourier decomposition as
\begin{align}
\phi(u,t, \vec{x}) = \int \frac{d \omega}{2 \pi} \frac{d^d k}{(2\pi )^2} ~e^{i \vec{k} \cdot \vec{x}} 
~T_{\omega}(x^{+})~ f_{\omega,\vec{k}}(u) ~\phi_0 (\omega,\vec{k}) \ ,
\label{IntTranAging}
\end{align}
where $\vec{k} $ is the momentum vector for the corresponding coordinates $\vec{x} $.
$\phi_0(\omega,\vec{k}) $ is introduced for the calculation of the correlation functions and 
is determined by the boundary condition with the normalization $f_{\omega,\vec{k}}(u_B) =1$. 
And $T_{\omega}(x^{+})$ is the kernel of integral transformation that convert $\omega$ to $x^{+}$, 
which is necessary for our time dependent setup \cite{Hyun:2011qj}. 

With this Fourier mode, the differential equation (\ref{BlukScalarEqAgingALCF}) decomposes into 
time dependent part and radial coordinate dependent one. 
The time dependent equation and solution read    
\begin{align}   \label{AgingTimeSol}
\left( i \frac{\partial}{\partial t} + \frac{\alpha M}{2 t} \right) T_{\omega}   
= \omega T_{\omega} \qquad \longrightarrow \qquad
T_{\omega}(t) 
=  c_1  \exp^{-i \omega t} t^{\frac{i\alpha M}{2}} \;.
\end{align}
The radial dependent equation is given by
\begin{align}
&u^2 f_{\omega,\vec{k}} '' +(z i M\alpha -d-1) u f_{\omega,\vec{k}}' 
-\left ( \frac{(d+2) z i M\alpha}{2} +\frac{z^2 \alpha^2 M^2}{4} + m^2 L^2 + \frac{\gamma M^2}{ u^{2z-4}}  \right) f_{\omega,\vec{k}} \nonumber \\
&= \left( \vec k^2  + 2 M\omega \right)~u^2 f_{\omega,\vec{k}} \;, 
\label{fEQ}
\end{align}
where $f^\prime=\partial_u f$. 

From this point we can not carry on the analysis for both the aging in ALCF, $\gamma=0$, and 
aging background $\gamma \neq 0$ simultaneously.  
Thus, we first consider the correlation functions of the scalar operator 
in aging ALCF.

\subsection{Aging in ALCF}    \label{sec:agingALCF}

For $\gamma=0$, an analytic solution of the equation (\ref{fEQ}) is available as  
\begin{align}
	f_{\omega,\vec{k}} =u^{\frac{2+d}{2}-i \frac{z \alpha M}{2}} 
	\left(  c_2 I_\nu (q u) + c_3  K_\nu (q u) \right) \;,
\end{align}
where 
$I_\nu$ and $K_\nu$ are Bessel functions with 
$\nu= \pm \sqrt{\left(\frac{d+2}{2}\right)^2+L^2 m^2 } $ and 
$q =   \sqrt{\vec k^2 +2 M \omega}$. Note the overall $\alpha$ dependent factor, 
which is a non-trivial feature of our model. 
We also consider the boundary condition along time direction near the boundary. 
The solution behaves as 
$f_q \sim f_{\omega,\vec{k}} \sim c_2 q^{-\nu} $ along with the time dependent factor 
$T_\omega$ in (\ref{AgingTimeSol}), whose inverse Fourier transform is given by 
\begin{align}
	\phi (x) \sim t^{\frac{\nu}{2}-\frac{d+2}{2}-\frac{\alpha \mathcal M}{2}} 
	\exp \left(- \frac{\mathcal M \vec x^2}{2 t} \right) \;.  
\end{align}
This wave function converges for $t \rightarrow \infty$ if 
$\frac{\nu}{2}-\frac{d+2}{2}-\frac{\alpha \mathcal M}{2} \leq  0$,
and for  $t \rightarrow 0$ due to the exponential factor if $\mathcal M >0 $. 
In particular, this condition allows the parameter $\alpha \mathcal M$ to be negative  
\begin{align}   \label{ConditionOnAlphaM}
	\alpha \mathcal M \geq \nu-d-2  \;, 
\end{align}
especially for the case $\nu <0$.
Similar result for $z=2$ and $d=2$ is already considered in \cite{Hyun:2011qj}. 
Note that we only consider the imaginary $M= i \mathcal M$. 

We follow \cite{Son:2002sd} to compute the correlation functions 
by introducing a cutoff $u_B$ near the boundary and 
normalizing $f_{\omega,\vec{k}}(u_B)  = 1$, which fixes 
$ c =  u_B^{-\frac{d+2}{2}+ i \frac{z \alpha M}{2}} K_\nu^{-1} (q u_B)$. 
The on-shell action is given by 
\begin{align}
	S[\phi_0] 
	&= \int d^{d+1} x  \frac{L^{d+3}}{u^{d+3}}  ~\phi^* (u,t, \vec{x}) ~\left(\frac{u^2}{L^2}\partial_u 
	+ i M \frac{z \alpha u}{2 L^2}  \right) \phi (u,t, \vec{x}) \Big |_{u_B} \;. 
\end{align}
This can be recast using 
\begin{align}
\phi(u,t, \vec y) = \int \frac{d^d k}{(2\pi )^d}\frac{d \omega}{2 \pi} e^{i k \cdot x} u^{\frac{d+2}{2}}  \left( \frac{t}{u^z} \right)^{\frac{i\alpha M}{2}} ~c~ K_\nu (q u) ~ \phi_0 (k) \;. 
\label{AgingWaveFunction}
\end{align}
as 
\begin{align}	
	&\int d t  ~\theta (t) ~\frac{d \omega'}{2 \pi} \frac{d \omega}{2 \pi} e^{-i (\omega' -\omega) t} 
	\left( t\right)^{-\frac{i\alpha (M^*-M)}{2}} 
	\nonumber \\
	&\qquad \times 
	\int d^2 y  \int \frac{d^d k'}{(2\pi)^d} \int \frac{d^d k}{(2\pi)^d} 
	e^{i(\vec k' - \vec k) \cdot \vec{x}} ~ \phi_0^* (k') \mathcal F(u, k', k) \phi_0 (k) \big |_{u_B} \;, 
	\label{wholeEQAging}
\end{align}
where $\theta (t)$ represents the existence of a physical boundary in the time direction, 
$0 \leq t < \infty$, and $\mathcal F$ is 
\begin{align}
	\mathcal F (u, k', k)
	&= \frac{L^{d+3}}{u^{d+3}} f_{k'}^* (k',u) \left(\frac{u^2}{L^2}\partial_u
	+ i M \frac{z \alpha u}{2 L^2}  \right) f_{k} (k,u) .
\end{align}
Note that the spatial integration along $\vec{x}$ can be done trivially to give 
a delta function $\delta^2 (\vec k' -\vec k)$. One can bring 
$u^{\pm i \frac{z \alpha M}{2}}$ factors in $f$ and $f^*$ together to cancel each other.
This removes the second part in $ \mathcal F$. From this point it is straight forward 
to check that $\mathcal F$ is given by (\ref{F1Function}) at the boundary. 
Further details can be found in \cite{Hyun:2011qj}.

For imaginary parameter $M = i \mathcal M$, we get the same correlation function as 
\begin{align}   \label{FiCorrelatorAgingLC}
&\langle \phi^* (x_{2}) \phi (x_{1}) \rangle =  
-2 \theta (t_2)~\left( \alpha^2 t_1 t_2 \right)^{\frac{\alpha \mathcal M}{2}} 
~(\alpha t_2)^{- \alpha \mathcal M}  \int \frac{d^{d+1} k}{(2\pi )^{d+1}} 
 e^{-i \vec{k} \cdot (\vec{x}_{1}-\vec{x}_{2})} e^{ i \omega (t_{1} - t_2) }  \mathcal F(u_{B},k)  \nonumber \\
 &=    \frac{\Gamma (1-\nu)}{ \Gamma (\nu) \Gamma (-\nu)}  
\frac{L^{d+1} \mathcal M^{\frac{d}{2}+\nu}  }{\pi 2^{\nu-1}  u_B^{d+2-2\nu}} 
\cdot \frac{\theta (t_2) \theta(t_2 - t_1) }{ (t_2 - t_1)^{\frac{d+2}{2}+\nu}}   
\cdot \left( \frac{t_2}{t_1} \right)^{-\frac{\alpha \mathcal M}{2} }   \cdot 
\exp \left( -{ \frac{\mathcal M (\vec{x}_2 - \vec{x}_1)^2 }{2 (t_2 - t_1)}}\right)  \;. 
\end{align}
This is one of our main results. The aging correlation functions for general dynamical exponent 
$z$ and $d$ have a direct relation with those of Schr\"odinger as  
\begin{align}
	\langle \phi^* (x_{2}) \phi (x_{1}) \rangle _{AgingALCF}^{z,d} =  
	\left( \frac{t_2}{t_1} \right)^{-\frac{\alpha \mathcal M}{2}} \langle \phi^* (x_{2}) \phi (x_{1}) \rangle _{ALCF}^{z,d} \;.
\end{align}
The overall time dependent factor $ \left( \frac{t_2}{t_1} \right)^{-\frac{\alpha \mathcal M}{2}}$ 
comes from the inverse Fourier transform of the time part, which 
has been evaluated in great detail \cite{Hyun:2011qj}. 
Thus the result is independent of $z$, while depending on the number of dimensions $d$.  
The corresponding extensions with log and log$^2$ are considered below in 
\S \ref{sec:AgingLogExtension}	.

\subsection{Aging backgrounds}    \label{sec:agingBack}

For aging backgrounds, we have nontrivial $z$ dependence, and we need to treat them separately. 
Fortunately, an analytic solution is available for $z=3/2$ 
\begin{align}   \label{Aging32Sol}
f_{\omega,\vec{k}} &= e^{-q u} u^{\nu+ \frac{2+d}{2}-i \frac{z \alpha M}{2}}   
\left( c_1  U [a,1+2\nu,2 qu ]
	+c_2  L_a^{2\nu} (2 q u) \right) \;, 
\end{align}
where $a=\frac{M^2}{2 q} +\frac{1+2\nu}{2} $, $\nu = \sqrt{(1+d/2)^2+ L^2 m^2}$ and 
$ q^2 = \vec k^2 + 2M\omega$. $U$ and $ L$ represent the
confluent hypergeometric function and the generalized Laguerre polynomial. 
We choose $U$ for our regular solution. 

The momentum space correlation function turns out to be the same as (\ref{SchrMomentumCorr}) 
as explained there. For the rest, we follow similarly \S \ref{sec:agingALCF} to get the 
aging correlation and response functions. Finally, we arrive general conclusion 
	\begin{align}   \label{AgingSchrCorrF1}
		&\langle \phi (x_{2}) \phi (x_{1}) \rangle_{Aging} 
		= \left( \frac{t_2}{t_1} \right)^{-\frac{\alpha \mathcal M}{2}}
		\langle \phi (x_{2}) \phi (x_{1}) \rangle_{Schr} \;.
	\end{align} 
The overall time dependent factor $ \left( t_2 / t_1 \right)^{-\frac{\alpha \mathcal M}{2}}$ 
comes from the inverse Fourier transform of the time part, which 
has been evaluated in great detail \cite{Hyun:2011qj}.

\subsection{Aging Response functions with log \& log$^2$ extensions}  \label{sec:AgingLogExtension}	

As we mentioned in \S \ref{sec:NTLFTsummary}, the aging in ALCF and aging background have 
similar properties as far as the correlation and response functions are concerned. 
Thus we present the logarithmic extension of the aging correlation functions 
using (\ref{FiCorrelatorAgingLC}). 

In the previous sections, \S \ref{sec:agingALCF} and \S \ref{sec:agingBack}, we establish the fact that 
the correlation functions have the overall time dependent factor 
$ \left( t_2 / t_1 \right)^{-\frac{\alpha \mathcal M}{2}}$ 
from the time dependent part of the momentum correlation function. 
In section \S \ref{sec:NTLFT}, on the other hand, we developed the algorithm to generate 
the logarithmic extensions using $\nu$ derivatives from the fact 
$ \left[ \mathrm  D \;, d/d \nu \right]  = 2 \nu$. These two generalizations are 
independent of each other. Thus we safely generate the logarithmic extensions of the 
aging correlation functions by differentiating the aging correlation functions in terms of $\nu$.  
	\begin{align}  \label{GeneralFormulaLogAging1}
		&\langle \phi (x_{2}) \phi (x_{1}) \rangle_{Aging}^{\mathcal F_1} 
		= \left( \frac{t_2}{t_1} \right)^{-\frac{\alpha \mathcal M}{2}}
		\langle \phi (x_{2}) \phi (x_{1}) \rangle_{Schr}^{\mathcal F_1}  \;, \\
		&\langle \phi (x_{2}) \phi (x_{1}) \rangle_{Aging}^{\mathcal F_2}  
		= \left( \frac{t_2}{t_1} \right)^{-\frac{\alpha \mathcal M}{2}} \frac{1}{2\nu} \frac{\partial}{\partial \nu} \left[ \langle \phi (x_{2}) \phi (x_{1}) \rangle_{Schr}^{\mathcal F_1}  
		\right]\;, \label{GeneralFormulaLogAging2} \\
		&\langle \phi (x_{2}) \phi (x_{1}) \rangle_{Aging}^{\mathcal F_3}  
		= \left( \frac{t_2}{t_1} \right)^{-\frac{\alpha \mathcal M}{2}}
		\frac{1}{4\nu} \frac{\partial}{\partial \nu} \left[ 
		\frac{1}{2\nu} \frac{\partial}{\partial \nu} \left[ 
		\langle \phi (x_{2}) \phi (x_{1}) \rangle_{Schr}^{\mathcal F_1}  
		\right] \right]  \;.	\label{GeneralFormulaLogAging3}	
	\end{align} 
The results $\langle \phi_2 \phi _1 \rangle^{\mathcal F_1}$ are given in 
equations (\ref{F1Correlator}), (\ref{SchrCorrF1}) and (\ref{SchrCorrLargeMF1}). 
Their specific forms are  
	\begin{align}  \label{F2CorrelatorAging}
		\langle \phi (x_{2}) \phi (x_{1}) \rangle_{Aging}^{\mathcal F_2}
		 &= \frac{1}{2 \nu^2} \left( 1 -\nu \psi (\nu) 
		+ \nu \ln [\mathcal M_{\Delta t}]
		 \right)  \langle \phi (x_{2}) \phi (x_{1}) \rangle_{Aging}^{\mathcal F_1} \;, 	\\	 
		\langle \phi_3 (x_{2}) \phi_3 (x_{1}) \rangle^{\mathcal F_3}_{Aging} 		
		\label{F3CorrelatorAging}		
		&= - \frac{1}{8 \nu^4} \left[ A_0 +A_1 \ln [\mathcal M_{\Delta t}] 
		+ A_2 \ln [\mathcal M_{\Delta t}]^2  \right]  
		\langle \phi (x_{2}) \phi (x_{1}) \rangle_{Aging}^{\mathcal F_1} \;, 
	\end{align}
where 
\begin{align}
	&\mathcal M_{\Delta t} = \frac{\mathcal M_B}{t_2 - t_1} 
	= \frac{\mathcal M u_B^2}{2(t_2 - t_1)}\;, \qquad 
	\psi(\nu) = \frac{\Gamma (\nu)'}{\Gamma (\nu)} \;, \\
	&A_0 =1 +\nu \psi - \nu^2( \psi^2- \psi')  \;, \quad 
	A_1 =( 2\nu^2 \psi - \nu) \;,  \quad  A_2 =- \nu^2 \;. 
\end{align}
These are main results of our aging response functions. 
Physical significances related to them are discussed in the following section.

\section{Connection to KPZ}	  \label{sec:KPZ}

In this section we would like to seek a connection to KPZ universality class, its growth,  
aging or both phenomena at the same time. 
Our investigation is concentrated on the generalizations of two-time response 
functions for general dynamical exponent $z$ and for general spatial dimensions $d$, along with 
their generalizations with the log and log$^2$ contributions.  

Previously, we observed that our two-time response functions reveal several qualitatively 
different behaviors, such as growth, aging or both in our holographic setup  \cite{Hyun:2012fd}. 
In a particular case, $z=2$ and $d=2$, both growing and aging behaviors have been observed 
for a parameter range $ -2 -\frac{\alpha \mathcal M}{2} <\nu < -2 $ \cite{Hyun:2012fd}. 
This was motivated by a recent progress on field theory side \cite{Henkel:2011NP}\cite{Henkel:2010hd} 
along with some clear experimental realization of the KPZ class in one spatial dimension 
\cite{TakeuchiPRL}\cite{TakeuchiSciRep}. 

Here we obtain additional properties of the two-time response 
functions as well as to extend our results for general $z$ and $d$. Before presenting the details, 
we comment their general behaviors. 
\begin{itemize}
\item[A.] Due to the simple broken time translation invariance of our system, 
signified by the parameter $\alpha \mathcal M$, 
our two-time response function reveals a power-law scaling behavior at early time region, 
which is distinct from another power-law scaling at late time region. 
The turning point between the two time regions, $y\approx 1$, is marked by the waiting time $s=t_1$. 
If $\alpha = 0$, there exists either only growth or aging behavior.  

\item[B.] The initial power scaling behaviors, growth or aging, are crucially related to 
the parameter $L^2 m^2$, especially the combination $-\frac{d+2}{2}-\nu $, 
which is the scaling dimensions of the dual field theory operators we consider.  
The late time scaling behavior is further modified by $\alpha \mathcal M $, 
which is aging parameter, in addition to the scaling dimensions. 

\item[C.] The power-law part of the two-time response functions show the growth and aging behaviors, 
while the log and log$^2$ corrections provides further modifications that would match 
detailed data by tuning available parameters.   

\end{itemize}

\subsection{Response functions for $z$ and $d$ : ALCF}

We consider a typical correlation and response functions for 
general $z$ and $d$, extending previous results for $z=2$ and $d=2$ \cite{Hyun:2012fd} 
	\begin{align}  \label{TwoTimeResponseTLCFTY}
		& C (s, y) = s^{-\frac{d+2}{2}-\nu} 
		y^{-\frac{\alpha \mathcal M}{2}-\frac{d+2}{2}-\nu} 
		\left(1 - \frac{1}{y} \right)^{-\frac{d+2}{2}-\nu} \;,
	\end{align}
with a waiting time, $s=t_1$, a scaling time, $y=t_2 / t_1$ and 
two other free parameters $\nu= \pm \sqrt{\left(\frac{2+d}{2}\right)^2+ L^2 m^2}$ 
and $\alpha \mathcal M$, which satisfies the condition (\ref{ConditionOnAlphaM}) 
coming from the time boundary $\alpha \mathcal M \geq \nu-d-2$. 
The response function (\ref{TwoTimeResponseTLCFTY}) is the general form  
for the Aging ALCF for all the cases considered in \S \ref{sec:agingALCF}. 
This is also valid for the ALCF in \S \ref{sec:ALCF} without the condition 
(\ref{ConditionOnAlphaM}) if we set $\alpha=0$.

\begin{figure}[!b]
\begin{center}
	 \includegraphics[width=0.47\textwidth]{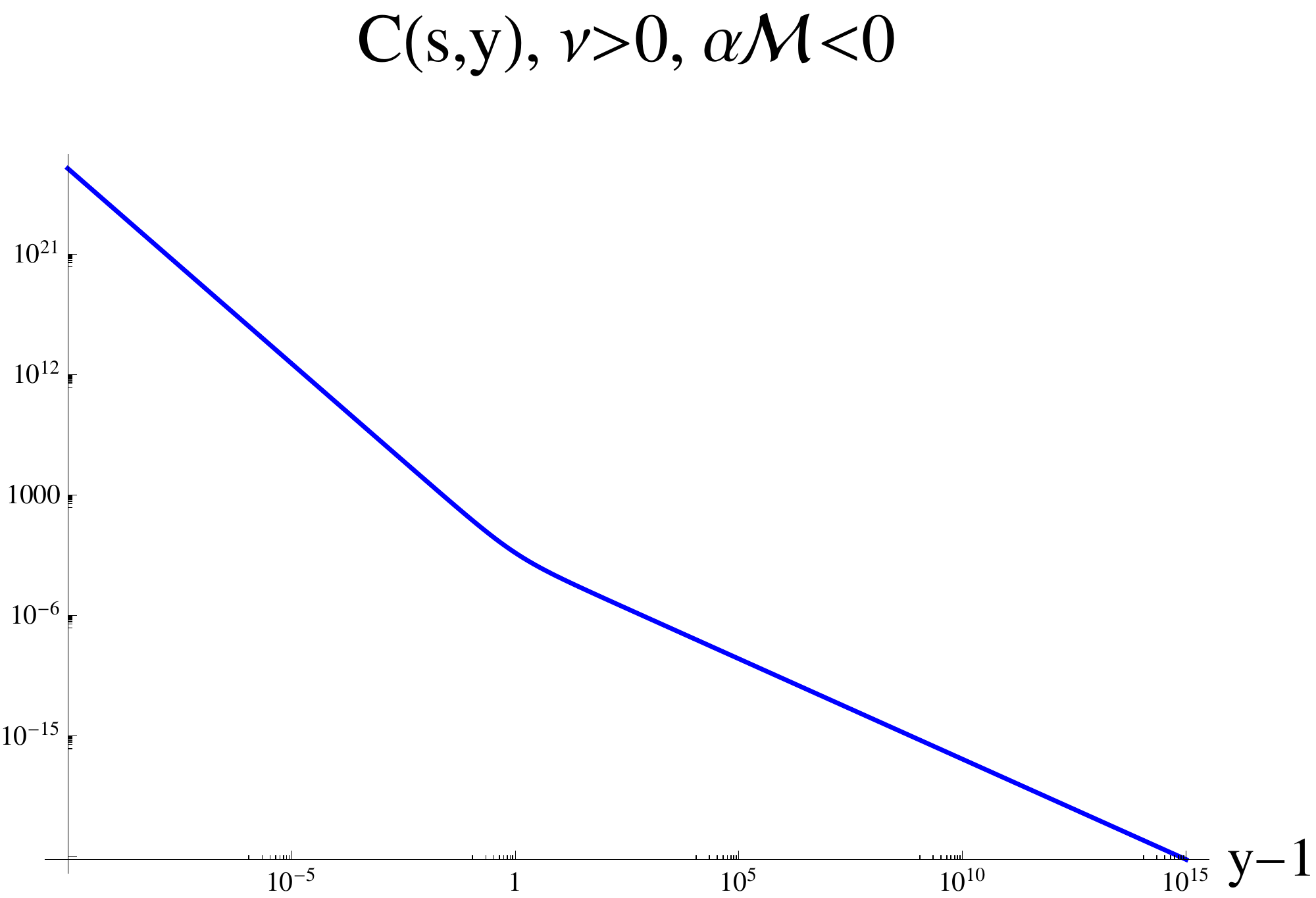}  \quad 
	 \includegraphics[width=0.47\textwidth]{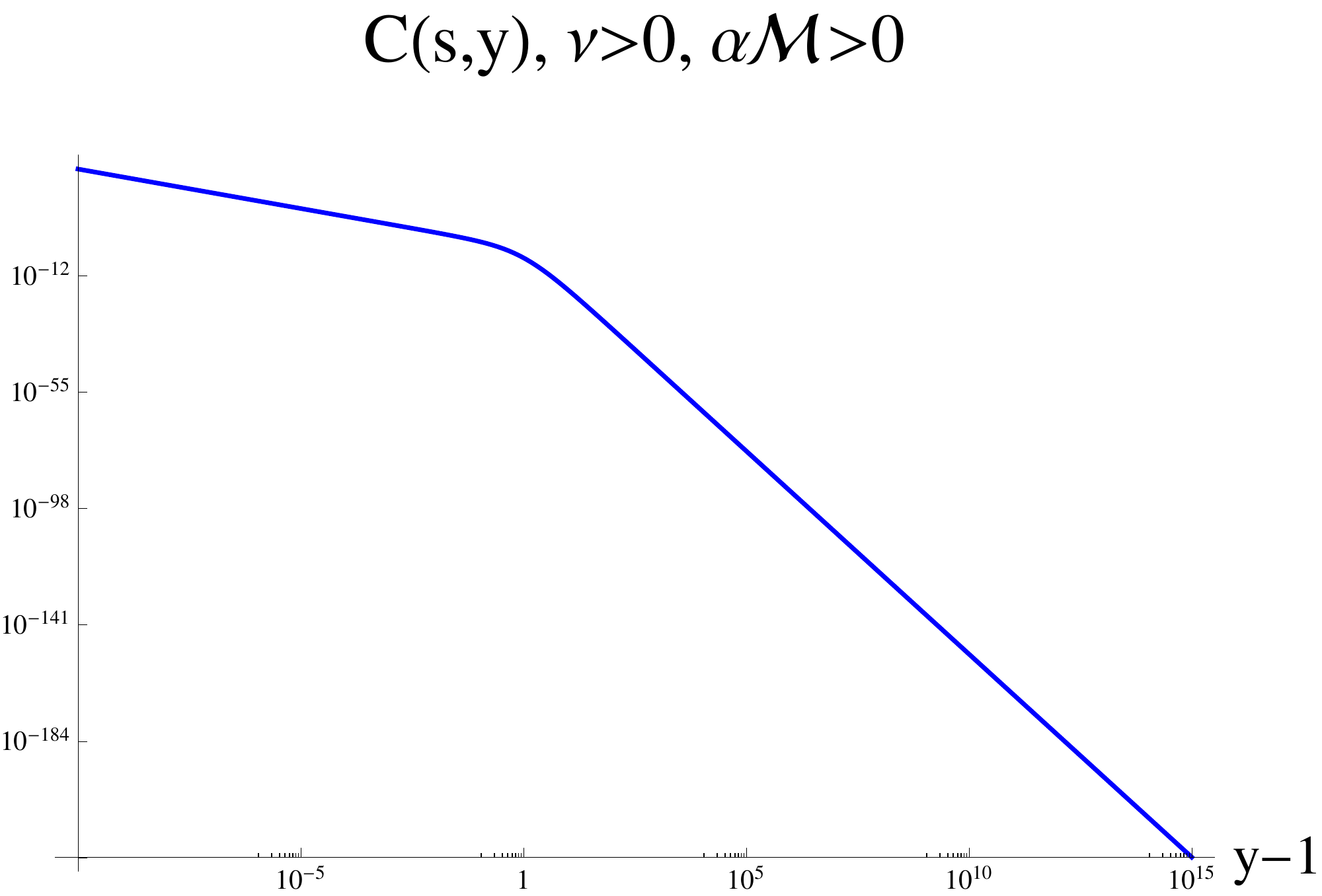}  
	 \caption{\footnotesize The log-log plots for the correlation functions with positive $\nu$. 
	 The parameters $d$ and $s$ do not change the qualitative behaviors, while the parameters $\nu$, 
	 actually $L^2 m^2$, and $\alpha \mathcal M$ are important for the early time and 
	 late time power law scaling.  }
	 \label{fig:PositiveNus}
\end{center}
\end{figure}

\bigskip 
{\it Positive $\nu$} 
 
Let us comment for the case with positive $\nu$. This case has only aging properties 
if the parameters $\nu$ and $\alpha \mathcal M$ are not too large.  
For $\nu \approx \frac{d+2}{2}$ and $ \alpha \mathcal M \approx \nu -d-2$, 
which is allowed by the time boundary condition (\ref{ConditionOnAlphaM}), 
the bending point, around $\log(y-1) \approx 1$ in the figure \ref{fig:PositiveNus}, 
sits deep down and the second leg of the plot becomes horizontal. 
As we increase either $\nu$ or $\alpha \mathcal M$, the bending point goes up. 
This is depicted in the figure \ref{fig:PositiveNus}. 
For $\alpha =0$ or a particular value of $L^2 m^2$, we can get a straight line, 
which is identical to the time independent case.  

\bigskip
{\it Negative $\nu$} 

\begin{figure}[!b]
\begin{center}
	 \includegraphics[width=0.49\textwidth]{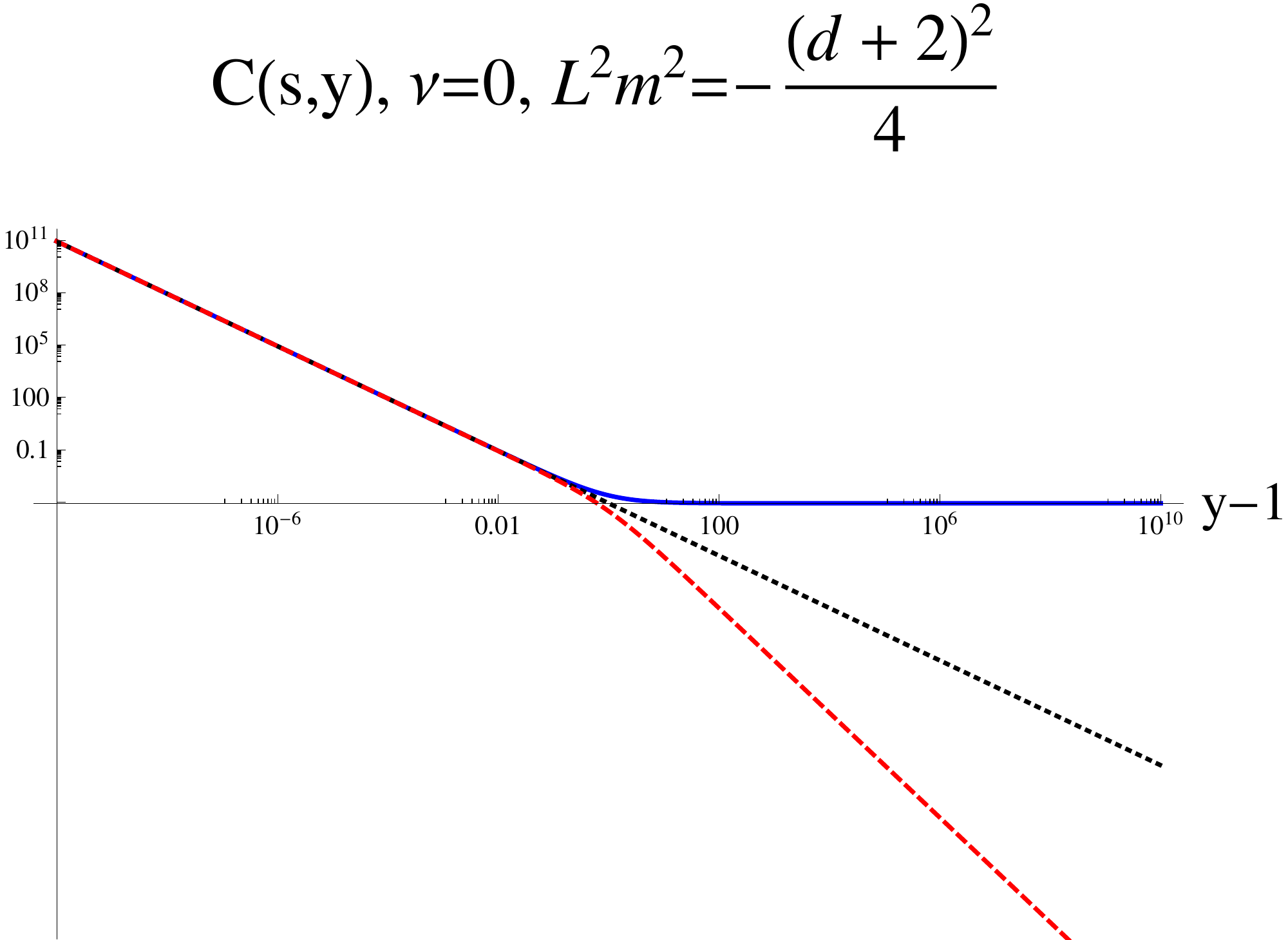}  
	 \includegraphics[width=0.49\textwidth]{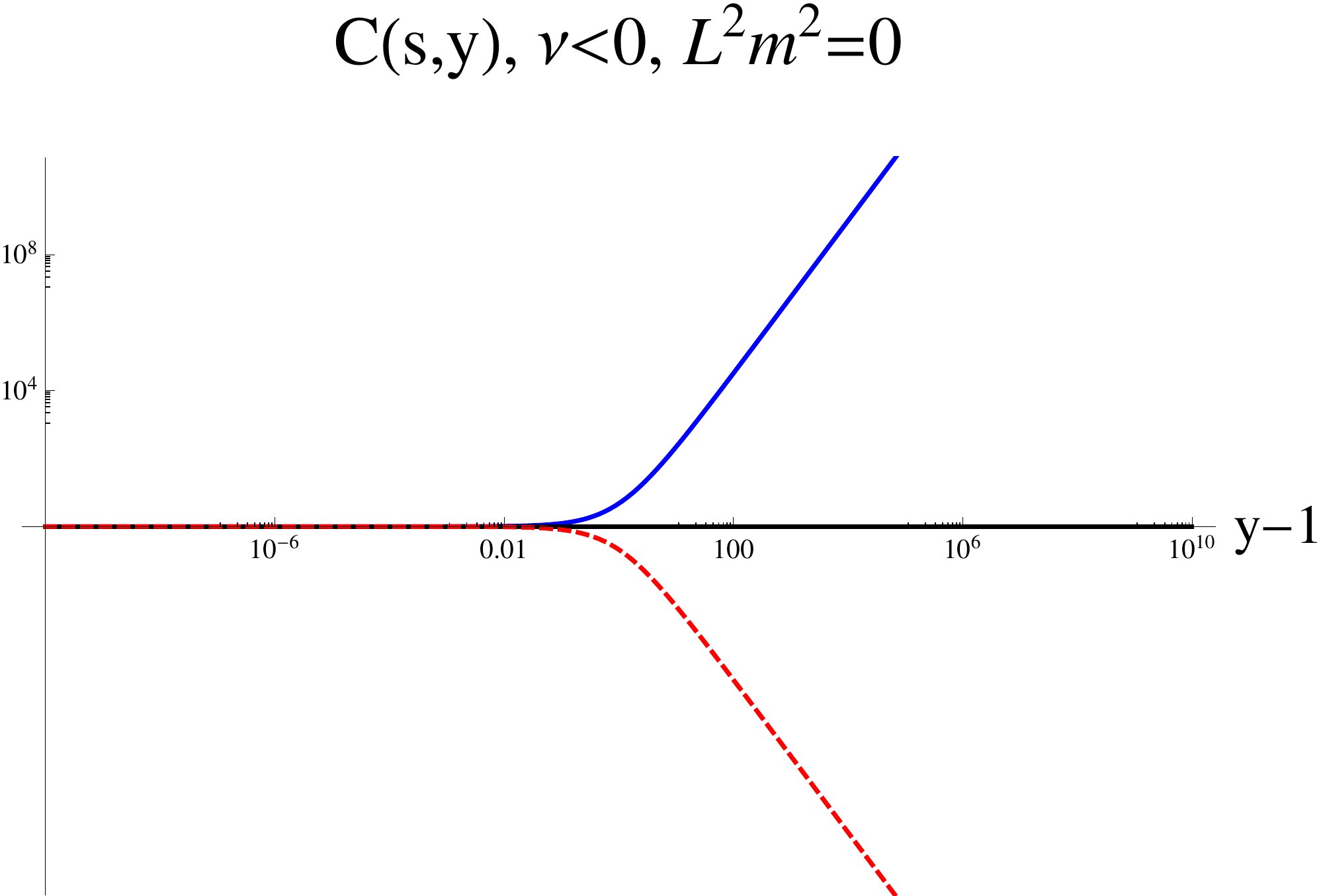}  \\
	 \includegraphics[width=0.49\textwidth]{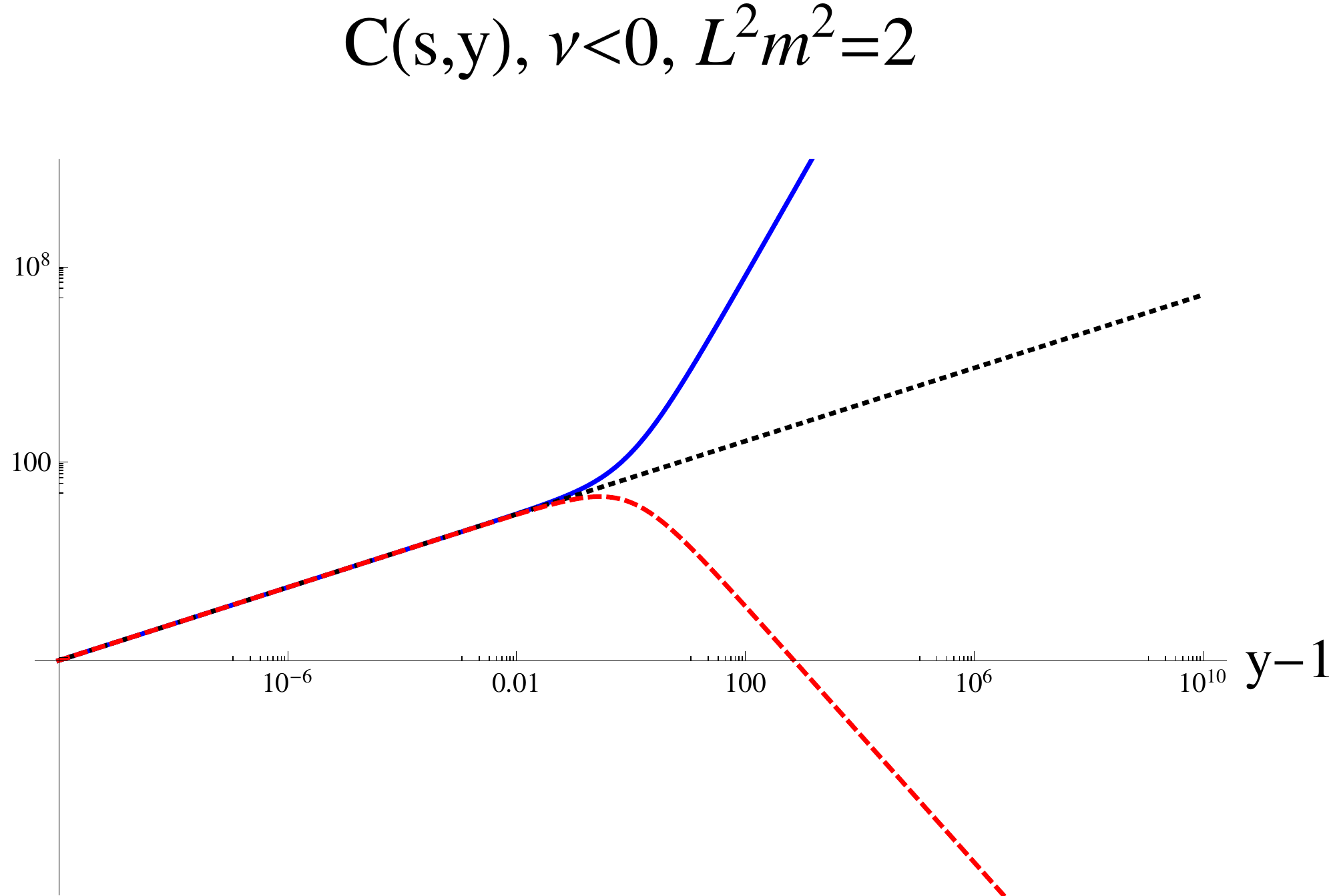}  
	 \caption{\footnotesize The log-log plots for the correlation functions with negative $\nu$. 
	 Each panel has a fixed value of $L^2 m^2$, which determines the slope of the first leg
	 at early time, while the sign of $\alpha \mathcal M$ determines the relative slope 
	 of the second leg at late time, compared to the first leg. 
	 Left : the case saturated with the BF bound $\nu=0$ and $L^2 m^2 = -(d+2)^2/4 $. 
	 Three plots are for $\alpha \mathcal M = \nu-d-2 < 0, 0, d+2-\nu>0 $, which are blue straight, 
	 black dotted and red dashed lines, respectively. Similarly for the Middle with $L^2 m^2 =0$ and 
	 Right plots with $L^2 m^2 =2 $ with the same values of 
	 $\alpha \mathcal M = \nu-d-2 < 0, 0, d+2-\nu>0 $.   }
	 \label{fig:NegativeNus}
\end{center}
\end{figure}

If one is interested in growth phenomena, it is more interesting to consider $\nu<0$. 
Due to the form of the response function (\ref{TwoTimeResponseTLCFTY}), 
the part $\left(1 - 1/y \right)^{-\frac{d+2}{2}-\nu} $ determines the properties at 
early time $ y \ll 1$. For $ \nu = - \sqrt{\left( 1+d/2 \right)^2 + L^2 m^2 }$, 
actually $L^2 m^2 $ determines the slope at early time. 
For $ -\left( 1+d/2 \right)^2 < L^2 m^2 < 0$, the slope of the first leg is negative, 
while that is positive for $L^2 m^2 > 0$. 
This can be verified directly in the figure \ref{fig:NegativeNus}.  

On the other hand, the late time behavior is determined by a factor 
$y^{-\frac{\alpha \mathcal M}{2}-\frac{d+2}{2}-\nu} $. If $\alpha=0 $, the slope 
does not change. The relative slope of the second leg is determined by the 
sign of $\alpha \mathcal M$. This is verified in the figure \ref{fig:NegativeNus}. 
The real slope of the second leg is governed by the sign of 
$-\frac{\alpha \mathcal M}{2}-\frac{d+2}{2}-\nu$.  In particular, 
early time growth and late time aging happens for  
\begin{align} 
	L^2 m^2 > 0 \;, \quad \alpha \mathcal M > -d-2 + \sqrt{\left( d+2 \right)^2 + 4 L^2 m^2 } \;.
\end{align} 
The second condition is similar to our time boundary condition (\ref{ConditionOnAlphaM}), but 
not identical.

\subsection{Critical exponents for general $z$ and $d$ : ALCF}  \label{sec:ALCFCriticalExponent}

For growth phenomena, the roughness of interfaces is quantified by their inter-facial width, 
$w (l, t) \equiv  \langle \sqrt{\langle [ h(x,t) - \langle h\rangle_l ]^2 \rangle_l }$,
defined as the standard deviation of the interface height $h (x, t) $ over a length scale 
$l $ at time $t $ \cite{BSBook}. An equivalent way to describe the roughness is the height-difference 
correlation function $ C(l, t) \equiv \langle [ h(x+l,t) - h(x,t) ]^2 \rangle $. 
$ \langle \cdots \rangle_l $ and $ \langle \cdots \rangle $ denote the average over a segment
of length $ l $ and all over the interface and ensembles, respectively.
Both $w (l, t)$ and $ C(l, t)^{1/2} $ are common quantities for characterizing the roughness,
for which the so-called Family-Vicsek scaling
\cite{FVScaling} is expected to hold. The dynamical scaling property is
	\begin{align}
		 C(l, t)^{1/2} \sim t^{\mathrm b} F(l t^{-1/z}) \sim \left\{
		 \begin{array}{ccc}
			l^{\mathrm a} & \text{for} & l \ll l_* \\
			t^{\mathrm b} & \text{for} & l \gg l_*  \end{array}
		 \right. \;,
	\end{align}
with two characteristic exponents: the roughness exponent ${\mathrm a} $ and the growth exponent
${\mathrm b} $. The dynamical exponent is given by
$z= \frac{\mathrm a}{\mathrm b} $, and the cross over length scale is $ l_* \sim t^{1/z} $.
For an infinite system, the correlation function behaves as $C(l, t)^{1/2} \sim t^{\mathrm b} $
at some late time region $t\gg 1$.

From the two-time correlation function in equation (\ref{TwoTimeResponseTLCFTY}), 
we can get the growth exponent 
\begin{align}
	\mathrm b_{Aging} = -\frac{\alpha \mathcal M}{4}-\frac{d+2}{4}-\frac{\nu}{2} \;, 
\end{align} 
where $\nu= \pm \sqrt{\left(\frac{2+d}{2}\right)^2+ L^2 m^2} $. 
Note that the parameters satisfy the condition (\ref{ConditionOnAlphaM}) from the 
time boundary conditions. We notice that our system size is infinite, and thus 
it is not simple matter to obtain the corresponding roughness exponent. 
The dynamical exponent is not fixed in ALCF, even though the differential 
equation has $z$ dependence, which can be checked in (\ref{BlukScalarEqAgingALCF}).  

For KPZ universality class, there is a nontrivial scaling relation between the 
roughness exponent $\mathrm a$ and the dynamical exponent $z$, chapter 6 in \cite{BSBook} 
\begin{align}   \label{azRelationKPZ}
	\mathrm a + z = 2 \;.
\end{align} 
While this relation is remained to be checked in our holographic model, we assume it is valid 
to make contact with some field theoretical models. 
Using the relation $ z = \frac{\mathrm a}{\mathrm b}$, we get 
\begin{align}
	&\mathrm a_{Aging} = \frac{2 \mathrm b_{Aging}}{\mathrm b_{Aging} +1} \;, \quad 
z_{Aging} = \frac{2}{\mathrm b_{Aging} +1} \;.
\end{align}

Let us examine these critical exponents against the known case for $d=1$ 
\begin{align}    \label{KPZExponents}
z_{KPZ}=\frac{3}{2} \;, \quad \mathrm a_{KPZ} = \frac{1}{2} \;,  \quad 
\mathrm b_{KPZ} = \frac{1}{3} \;. 
\end{align} 
These can be reproduced with the condition 
\begin{align}
	\alpha \mathcal M + 2 \nu = -\frac{13}{3} \;, 
\end{align} 
which can be matched for $L^2 m^2 = 2.444 \cdots$ and $ \alpha \mathcal M = 0$ 
for negative $\nu$. We choose $ \alpha \mathcal M = 0$ for the simple growth behavior.

\subsubsection{Negative $\nu$}

There are two independent critical exponents. One particular interesting exponent is 
the so called growth exponent $\mathrm b = -\frac{\alpha \mathcal M}{4}-\frac{d+2}{4}-\frac{\nu}{2}$, 
where $\nu= \pm \sqrt{\left(\frac{2+d}{2}\right)^2+ L^2 m^2} $. We consider a dual field 
theory operator with $\nu= - \sqrt{\left(\frac{2+d}{2}\right)^2+ L^2 m^2} $, 
$L^2 m^2 \ll \left( \frac{2+d}{2} \right)^2 $.  
Then by expanding for small $m$, we get 
\begin{align}
	\mathrm b_{Aging} 
	\approx  \frac{L^2 m^2}{2(d+2)} -\frac{\alpha \mathcal M}{4} \;.
\end{align}

Using again $\mathrm a + z = 2 $ and the relation $ z = \frac{\mathrm a}{\mathrm b}$, we get 
\begin{align}
	&\mathrm a_{Aging} 	\approx \frac{\frac{L^2 m^2}{d+2} -\frac{\alpha \mathcal M}{2}}{1+\frac{L^2 m^2}{2(d+2)} -\frac{\alpha \mathcal M}{4}} \;, \quad
	z_{Aging} \approx \frac{2 }{1+\frac{L^2 m^2}{2(d+2)} -\frac{\alpha \mathcal M}{4}} \;. 
\end{align}

\begin{figure}[!ht]
\begin{center}
	 \includegraphics[width=0.48\textwidth]{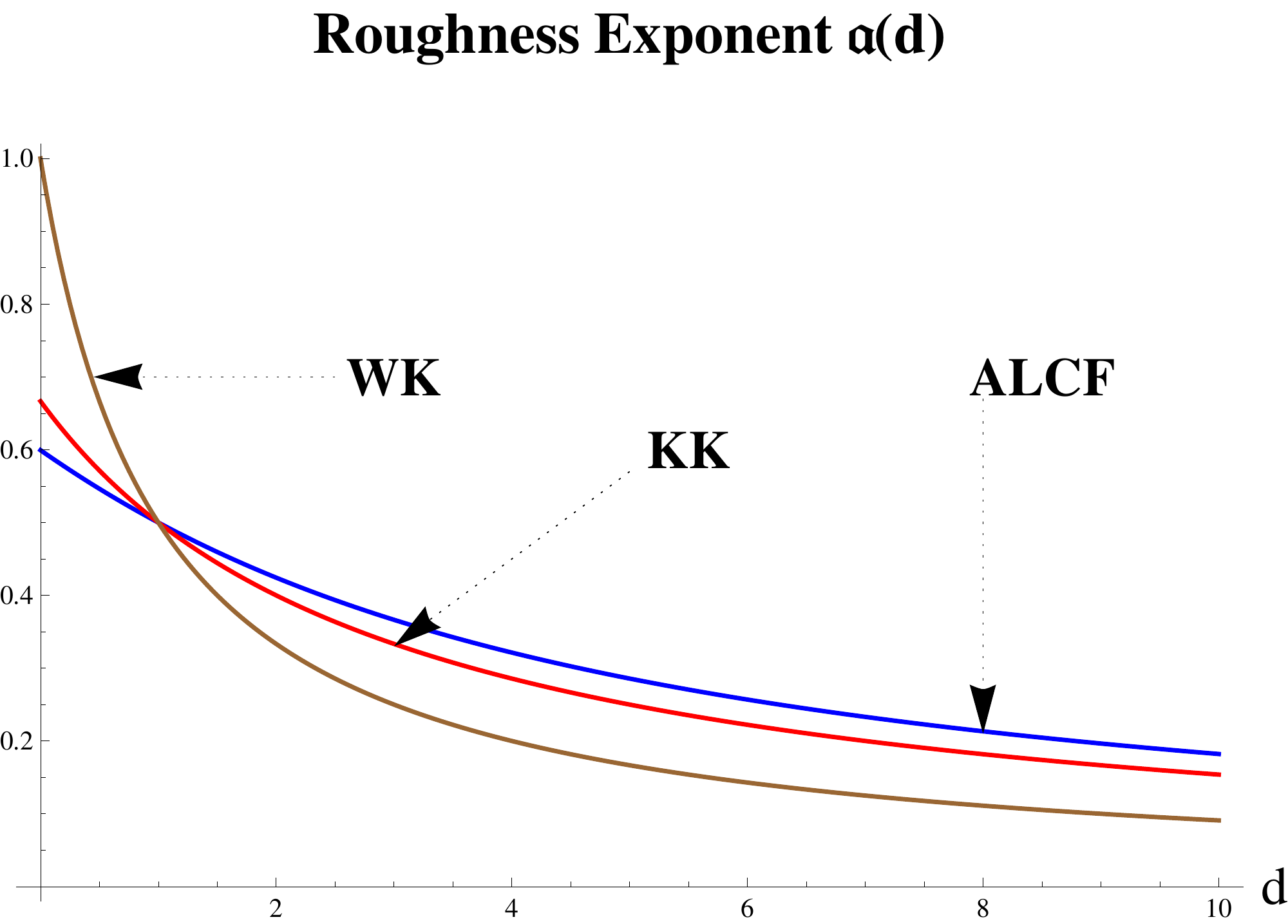}  \quad 
	 \includegraphics[width=0.48\textwidth]{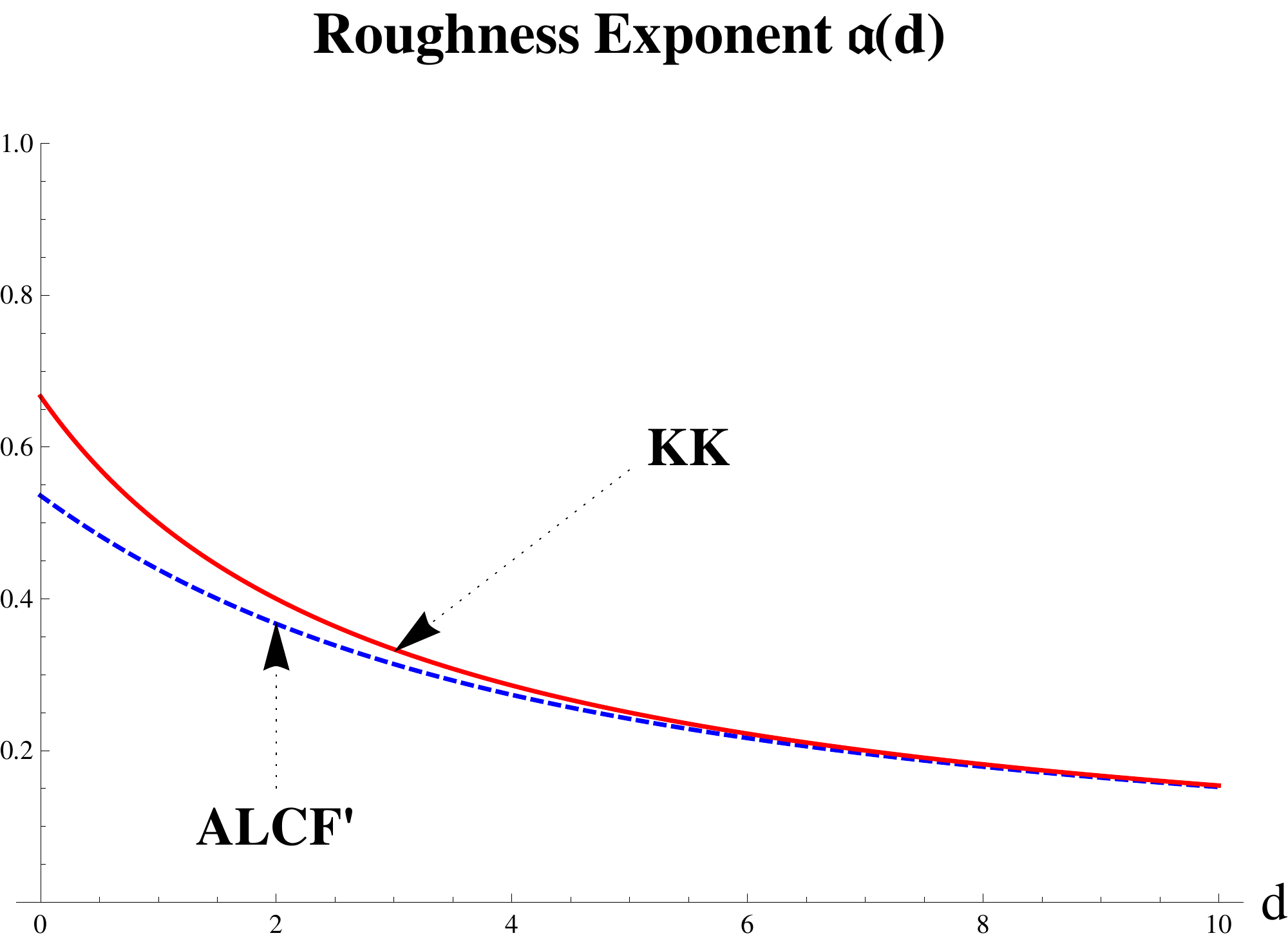}  
	 \caption{\footnotesize Left panel : plot for the roughness exponent $\mathrm a$ of ALCF for 
	 $\alpha \mathcal M = 0 $ and $L^2 m^2 \approx 2.444 \cdots $ that matches KPZ exponents for $d=1$ 
	 given in (\ref{KPZExponents}). 
	 KK represents $ \mathrm a_{KK} = \frac{2}{d+3}$ from Kim-Kosterlitz \cite{KimKosterlitz1989}, 
	 while WK $ \mathrm a_{WK} = \frac{1}{d+1}$ from Wolf-Kert\'esz \cite{WolfKertesz1987}. 
	 Right panel : plot for $\mathrm a$ of ALCF$'$ for 
	 $\alpha \mathcal M = 0 $ and $L^2 m^2 = 2$ that matches (\ref{KKExponents}) 
	 only when $L^2 m^2 = 2 \ll \left(\frac{2+d}{2}\right)^2 $. }
	 \label{fig:RoughnessExponents}
\end{center}
\end{figure}

If we further restrict our attention to the case $\alpha \mathcal M = 0$ for considering 
only the growth phenomena, we have the following dependence on the number of spatial dimensions 
\begin{align} 
	\mathrm b_{ALCF} \approx  \frac{Z}{d+2} \;, \qquad 
	\mathrm a_{ALCF} \approx \frac{2 Z}{d+2+ Z} \;, \qquad 
	z_{ALCF} \approx 2 \frac{d+2}{d+2+ Z} \;, 
\end{align} 
where $Z = \frac{L^2 m^2 }{2}$. 
For $Z=1.222 \cdots$, these exponents match (\ref{KPZExponents}) for $d=1$. 
The corresponding roughness exponent is depicted as blue line in the left panel of the figure 
\ref{fig:RoughnessExponents}, which is referred as ``ALCF''. 
To compare with other growth models \cite{BSBook}, we also depicted the roughness exponents of the 
Kim-Kosterlitz model $ \mathrm a_{KK} = \frac{2}{d+3}$ \cite{KimKosterlitz1989} as well as 
Wolf-Kert\'esz model $ \mathrm a_{WK} = \frac{1}{d+1}$ \cite{WolfKertesz1987}. 

For $Z=1$, we get 
\begin{align}   \label{KKExponents} 
	\mathrm b_{ALCF} \approx  \frac{1}{d+2} \;, \qquad 
	\mathrm a_{ALCF} \approx \frac{2}{d+3} \;, \qquad 
	z_{ALCF} \approx 2 \frac{d+2}{d+3} \;, 
\end{align} 
Our results (\ref{KKExponents}) are only valid for $L^2 m^2 = 2 \ll \left(\frac{2+d}{2}\right)^2 $, 
which is referred as ``ALCF$'$'' in the right panel of the figure \ref{fig:RoughnessExponents}.
These exponents have been conjectured for growth in a restricted solid-on-solid model 
by Kim-Kosterlitz \cite{KimKosterlitz1989}\cite{BSBook}.

\subsection{Response functions with log extensions : ALCF}

We have shown in previous sections that the response functions reveal growth and aging 
behaviors without log or log$^2$ corrections. The log and log$^2$ corrections have been  
considered to match further details at early time region \cite{Henkel:2011NP}\cite{Henkel:2010hd}. 
In this section, we would like to investigate some more details related to those corrections 
based on previous results \cite{Hyun:2012fd}.

\subsubsection{With log extension} 

Our two-time response functions with log correction are given by 
	\begin{align}  \label{TwoTimeResponseTLCFTYlog}
		C_{log}(s, y) &= s^{-\frac{d+2}{2}-\nu} 
		y^{-\frac{\alpha \mathcal M}{2}-\frac{d+2}{2}-\nu} 
		\left(1 - \frac{1}{y} \right)^{-\frac{d+2}{2}-\nu} \nonumber \\
		&\quad \times \bigg\{1 + R_1 (\ln s + \ln y) 
		+ R_1 \ln [1 - \frac{1}{y}]  \bigg\} \;,
	\end{align}
where $s=t_1$, $y=\frac{t_2}{t_1}$, $\nu$ and $\alpha \mathcal M$ are free parameters, 
while the coefficients are given by 
\begin{align} 
&R_1 = \frac{\nu}{1- \nu \psi(\nu) + \nu \ln [\mathcal M_B]} \;.
\end{align} 
We note that the coefficients are completely fixed by two fixed parameters 
\begin{align}  \label{MBPsiParameters}
&\mathcal M_B = \frac{\mathcal M u_B^2}{2} \;, \qquad  
\psi(\nu) = \frac{\Gamma (\nu)'}{\Gamma (\nu)} \;.
\end{align} 
Note that similar result for $z=2$ and $d=2$ has been available in \cite{Hyun:2012fd}. 
The detailed comparisons between (\ref{TwoTimeResponseTLCFTYlog}) and the phenomenological 
field theory model \cite{Henkel:2011NP}\cite{Henkel:2010hd} were investigated. 
We noted that the terms proportional to $\ln s $ and $ \ln y$ are not considered in 
\cite{Henkel:2011NP}\cite{Henkel:2010hd}, which do not modify qualitative features of 
the response functions. For this case, the analysis done in \cite{Hyun:2012fd} is still valid.

\subsubsection{With log$^2$ extension} 

\begin{figure}[!ht]
\begin{center}
	 \includegraphics[width=0.47\textwidth]{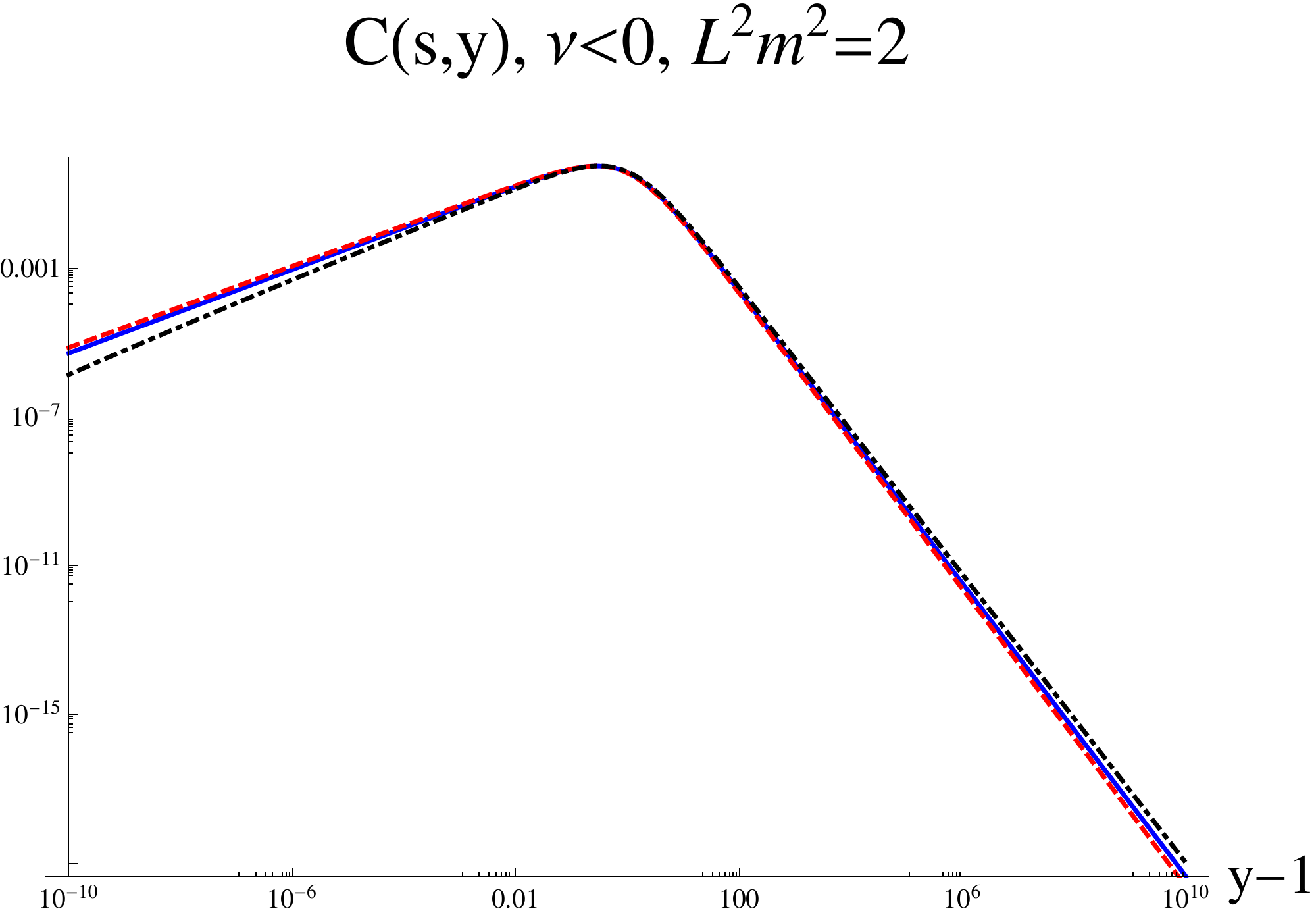}  \quad 
	 \includegraphics[width=0.47\textwidth]{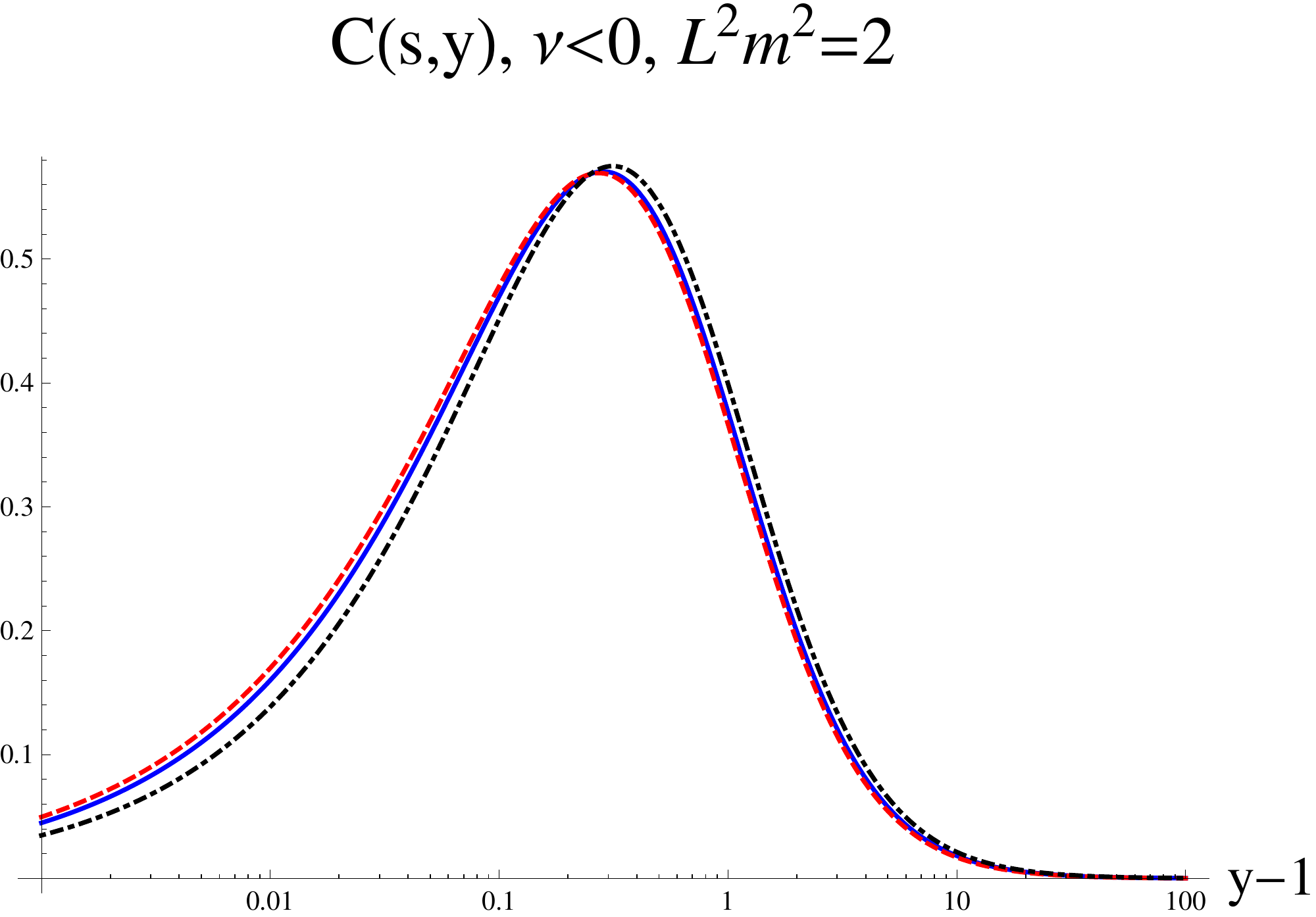}  
	 \caption{\footnotesize Plots for $ C(s, y)$ with blue straight, $ C_{log}(s, y)$ 
	 with red dashed and $ C_{log^2}(s, y)$ with black dot-dashed lines for $\nu <0$, $d=1$, $s=4$ and 
	 $L^2 m^2 =2$. For the response functions with log corrections, we need one more input 
	 $\mathcal M_B$, which we took $\mathcal M_B = 10^{-15}$ for these plots. 
	 The smaller the value of $\mathcal M_B$, the smaller the 
	 differences between $ C(s, y)$ and its log extensions.  }
	 \label{fig:LogNegativeNus}
\end{center}
\end{figure}

We obtain the log$^2$ extension of the response function using holographic approach 
	\begin{align}  \label{TwoTimeResponseTLCFTYlog2}
		& C_{log^2}(s, y) = s^{-\frac{d+2}{2}-\nu} 
		y^{-\frac{\alpha \mathcal M}{2}-\frac{d+2}{2}-\nu} 
		\left(1 - \frac{1}{y} \right)^{-\frac{d+2}{2}-\nu}  \nonumber \\
		&\qquad \qquad \quad \times \bigg\{1 + \tilde R_1 (\ln s + \ln y) + \tilde R_2 (\ln s + \ln y)^2 \bigg. \nonumber \\
		&\qquad \qquad \qquad  \bigg. + [ \tilde R_1 + 2 \tilde R_2 (\ln s + \ln y) ]  \ln [1 - \frac{1}{y}] + \tilde R_2 \ln [1 - \frac{1}{y}]^2  \bigg\} \;, 
	\end{align}
where 
\begin{align}    \label{Log2Coefficients}
&\tilde R_1 = - \frac{\tilde A_1 + 2 \tilde A_2 \ln [\mathcal M_B]}{\tilde A_0 + \tilde A_1 \ln [\mathcal M_B] + \tilde A_2 \ln [\mathcal M_B]^2}  \;,  \\
&\tilde R_2 = \frac{\tilde A_2}{\tilde A_0 + \tilde A_1 \ln [\mathcal M_B] + \tilde A_2 \ln [\mathcal M_B]^2} \;, \\
&\tilde A_0 =1 +\nu \psi - \nu^2( \psi^2- \psi') \;, \\
&\tilde A_1 =( 2\nu^2 \psi - \nu) \;, \qquad \tilde A_2 =- \nu^2  \;.
\end{align} 
These coefficients are also completely fixed by two fixed parameters 
given in (\ref{MBPsiParameters}).  
Compared to $C(s,y)$, a new parameter $\mathcal M_B$ determines the behaviors of 
response functions related to the log and log$^2$ contributions. As we explicitly check in 
the figure \ref{fig:LogNegativeNus}, the qualitative behavior of the response functions 
does not change with the log and log$^2$ contributions for reasonably small $\mathcal M_B$.  
The growth exponent and aging properties are determined by the two parameters 
$L^2 m^2$ and $\alpha \mathcal M$, which define our theory. 

\begin{figure}[!hb]
\begin{center}
	 \includegraphics[width=0.47\textwidth]{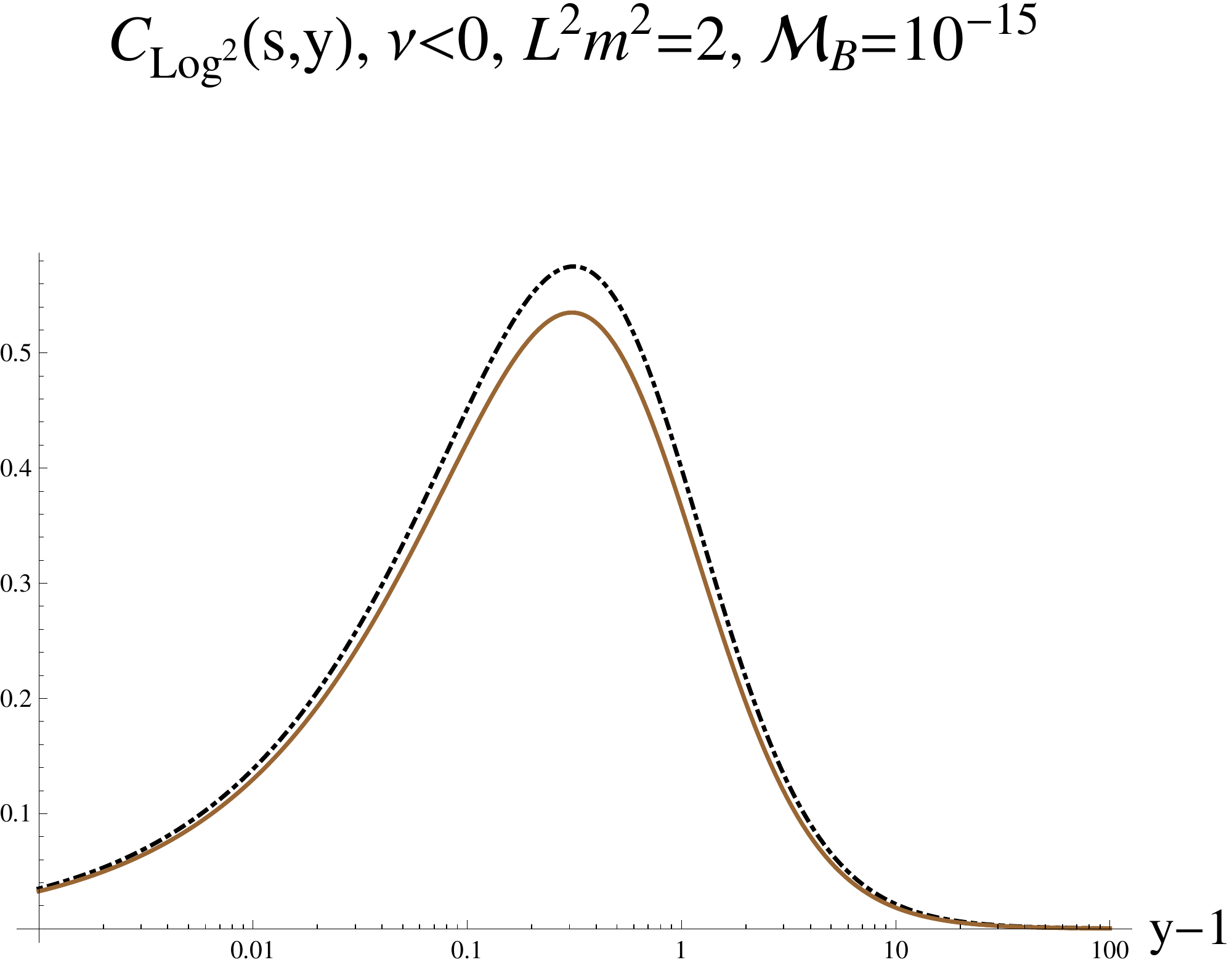}  \quad 
	 \includegraphics[width=0.47\textwidth]{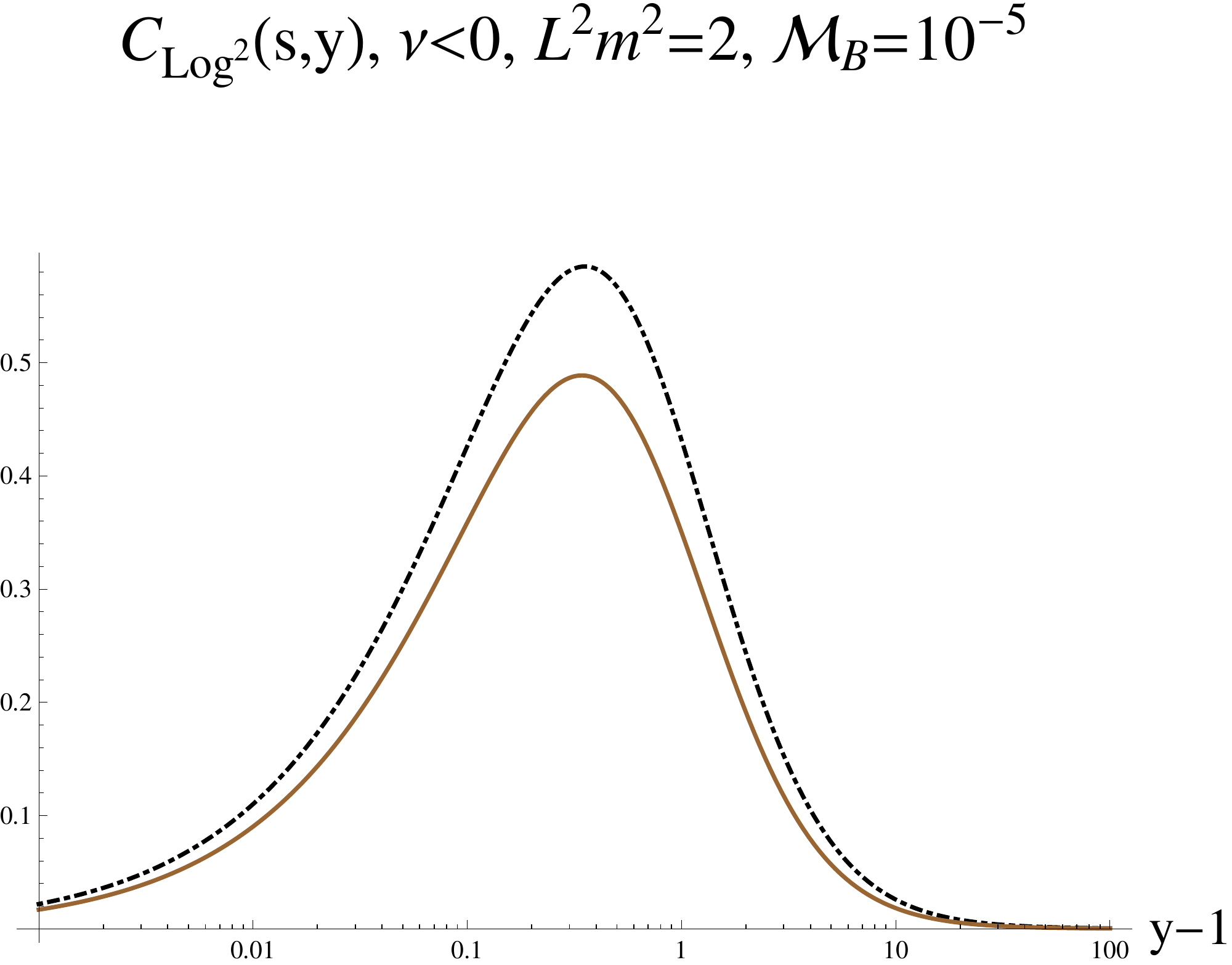}  
	 \caption{\footnotesize Plots for $ C_{Log^2}(s, y)$ with black dot-dashed and 
	 $ C_{Log^2}^{Wanted}(s, y)$ with brown straight lines for $\nu <0$, $d=1$, $s=4$ and 
	 $L^2 m^2 =2$. Left : $\mathcal M_B = 10^{-15}$, Right : $\mathcal M_B = 10^{-5}$. 
	 We check that there is no qualitative changes due to the unwanted terms explained in 
	 (\ref{TwoTimeResponseTLCFTYlog2Wanted}). }
	 \label{fig:LogUnWantedTerms}
\end{center}
\end{figure}

We would like to compare our results (\ref{TwoTimeResponseTLCFTYlog2}) with the following equation, 
which is equation (10)  of \cite{Henkel:2011NP} (or equation (4.3) of \cite{Henkel:2010hd}), 
obtained from the phenomenological field theory model.
	\begin{align}   \label{FTresponseFunction}
		R(s,y)&=s^{-1-a} y^{-\lambda_R/z} \left(1-\frac{1}{y}\right)^{-1-a'}  \\
		&\times\left(h_0-g_{12,0}\tilde\xi'\ln [ 1- \frac{1}{y}]-\frac{1}{2} f_{0}\tilde \xi'^2\ln^2 [1-\frac{1}{y}]-g_{21,0}\xi'\ln[y-1]+\frac{1}{2}f_{0}\xi'^2\ln^2 [y-1]\right) \;, \nonumber
	\end{align}
where the parameters $g_{12,0},g_{21,0},\tilde\xi',\xi',f_{0}$ come from the logarithmic extension
in the field theory side. From the phenomenological input \cite{Henkel:2010hd}
``the  parenthesis  becomes  essentially  constant  for  sufficiently  large $   y $,''
the condition $   \xi ' =0 $ is imposed to remove the last two terms.
The other parameters in (\ref{FTresponseFunction}) are determined to match available data.
We can identify the exponents $a, a', \lambda_R$ by comparing the equation to our result
\cite{Hyun:2012fd}
	\begin{align}
	\nu=a-1=a'-1 \;, \qquad \frac{\alpha\mathcal M}{2}=\frac{\lambda_R}{z}-1-a \;.
	\end{align}

By including the log$^2$ corrections, we provide the necessary terms 
$ \tilde R_2 \ln [1 - \frac{1}{y}]^2 $ as well as $ \tilde R_1 \ln [1 - \frac{1}{y}] $.  
The relative coefficients between them are fixed by (\ref{Log2Coefficients}). 
On the other hand, there exist also several unwanted terms inside the parenthesis of 
(\ref{TwoTimeResponseTLCFTYlog2}).
First, we note new terms $ \tilde R_2 (\ln s + \ln y)^2$ 
and $ 2 \tilde R_2 (\ln s + \ln y)   \ln [1 - \frac{1}{y}]	$ due to log$^2$ corrections, 
in addition to $ \tilde R_1 (\ln s + \ln y)$ coming from the log correction. 
All these terms might spoil the desired properties of the phenomenological 
response function (\ref{FTresponseFunction}). To examine the effects coming from these 
unwanted terms, we plot the response function with only wanted terms as 
	\begin{align}  \label{TwoTimeResponseTLCFTYlog2Wanted}
		C_{log^2}^{Wanted}(s, y) &= s^{-\frac{d+2}{2}-\nu} 
		y^{-\frac{\alpha \mathcal M}{2}-\frac{d+2}{2}-\nu} 
		\left(1 - \frac{1}{y} \right)^{-\frac{d+2}{2}-\nu} \nonumber \\
		&\quad \times \bigg\{1 + \tilde R_1 \ln [1 - \frac{1}{y}] 
		+ \tilde R_2 \ln [1 - \frac{1}{y}]^2  \bigg\} \;. 
	\end{align}
We explicitly check in the figure \ref{fig:LogUnWantedTerms} that 
the full response functions (\ref{TwoTimeResponseTLCFTYlog2}) 
have a qualitatively similar behavior compared to those (\ref{TwoTimeResponseTLCFTYlog2Wanted}) 
with only wanted terms in the field theory approach.

\subsection{Critical exponents for $z=\frac{3}{2}$ and $d=1$ : Aging backgrounds}

While the dynamical exponent $z$ for the ALCF is not fixed in obtaining the correlation and 
response functions, those of the Schr\"odinger backgrounds crucially depend on $z$. 
In fact, obtaining analytic solutions of the differential equation (\ref{BlukScalarEqSchr}) 
is a highly non-trivial task. Fortunately, we are able to get response functions for $z=\frac{3}{2}$ 
with some approximations as
	\begin{align}  \label{TwoTimeResponseTLCFTYSchr2}
		& C_{Schr} (s, y) = s^{-\frac{d+2}{2}-f \nu} 
		y^{-\frac{\alpha \mathcal M}{2}-\frac{d+2}{2}-f \nu} 
		\left(1 - \frac{1}{y} \right)^{-\frac{d+2}{2}-f \nu} \;,
	\end{align}
where $s=t_1$, $y=\frac{t_2}{t_1}$, $\nu$ and $\alpha \mathcal M$ are free parameters. 
This is valid for the Aging background (\ref{AgingSchrCorrF1}) as well as 
the Schr\"odinger background, $\alpha=0$, given by (\ref{SchrCorrF1}) with $f=1$ for 
$\mathcal M \rightarrow 0$ and by (\ref{SchrCorrLargeMF1}) with $f=2$ for 
$\mathcal M \rightarrow \infty$.

From the two-time response functions in equation (\ref{TwoTimeResponseTLCFTY}), 
we can get the growth exponent 
\begin{align}
	\mathrm b_{Aging}^{f} = -\frac{\alpha \mathcal M}{4}-\frac{d+2}{4}-\frac{f \nu}{2} \;, 
\end{align} 
where $\nu= \pm \sqrt{\left(\frac{2+d}{2}\right)^2+ L^2 m^2} $. 
Now the dynamical exponent is fixed as $z=\frac{3}{2}$ for the aging background. 
Using the relation $ z = \frac{\mathrm a}{\mathrm b}$,  
we get 
\begin{align}
	&\mathrm a^f_{Aging} = \frac{3 \mathrm b_{Aging}}{2}
	= -\frac{3\alpha \mathcal M}{8}-\frac{3(d+2)}{8}-\frac{3 f \nu}{4} \;.  
\end{align}

The critical exponents for the KPZ universality class given in (\ref{KPZExponents}) can 
be reproduced with the condition 
\begin{align}
	\alpha \mathcal M + 2 f \nu = -\frac{13}{3} \;, 
\end{align} 
which can be matched for $ \alpha \mathcal M \approx 0$ and 
$L^2 m^2 \approx \left(\frac{13}{6 f}\right)^2 - \frac{9}{4}$  
for negative $\nu$. The value of $L^2 m^2$ for $f=2$ becomes negative, yet is allowed 
as we can see from the expression of $\nu$.

\section{Conclusion}   \label{sec:conclusion}

We have extended our geometric realizations of aging symmetry in several different ways 
based on previous works for $z=2$ and $d=2$ \cite{Hyun:2012fd}\cite{Hyun:2011qj}. 
First, we generalize our correlation and response functions to the non-conformal setup 
with general dynamical exponent $z$ and for arbitrary spatial dimensions $d$. 
They have Galilean symmetries with time translation symmetry, which are summarized 
in the equations (\ref{F1Correlator}), (\ref{SchrCorrF1}) and (\ref{SchrCorrLargeMF1}). 
For convenience, we reproduce equation (\ref{F1Correlator}) here 
\begin{align}
\langle \phi^* (x_{2}) \phi (x_{1}) \rangle
= \frac{\Gamma (1-\nu)}{ \Gamma (\nu) \Gamma (-\nu)}  
\frac{L^{d+1} \mathcal M^{\frac{d}{2}+\nu}  }{\pi 2^{\nu-1}  u_B^{d+2-2\nu}} 
\cdot \frac{\theta (t_2) \theta(t_2 - t_1) }{ (t_2 - t_1)^{\frac{d+2}{2}+\nu}}   
\cdot \exp \left( -{ \frac{\mathcal M (\vec{x}_2 - \vec{x}_1)^2 }{2 ( t_2 - t_1)}}\right)  \;, 
\end{align}
which is valid for general $z$ and $d$. 
Second, these are extended with log and log$^2$ corrections with appropriate 
bulk actions. Practically, these corrections can be computed using simple properties of the 
differential operators (\ref{DiffOpALCF}), (\ref{DiffOpSchr}) and their commutation 
relations (\ref{CommutationRelation}). The results are listed in (\ref{F2Correlator}), 
(\ref{F3Correlator}) for ALCF, (\ref{SchrCorrF2}), (\ref{SchrCorrF3}) for 
Schr\"odinger backgrounds. All these response functions have time translation invariance. 

Third, on top of these extensions, we also compute response functions with aging symmetry, 
by breaking the global time translation invariance using a singular coordinate 
transformation (\ref{CoordinateChangez}). We check the general relation between the aging 
response functions and those of Schr\"odinger backgrounds holds 
(\ref{GeneralFormulaLogAging1})
\begin{align}
	\langle \phi^* (x_{2}) \phi (x_{1}) \rangle _{Aging}^{z,d} =  
	\left( \frac{t_2}{t_1} \right)^{-\frac{\alpha \mathcal M}{2}} 
	\langle \phi^* (x_{2}) \phi (x_{1}) \rangle_{Schr}^{z,d} \;.
\end{align}
This generalization is independent of the logarithmic extensions. 

With these results, we investigate our two-time response functions for general $z$ and 
especially for arbitrary number of spatial dimensions $d$ with log and log$^2$ extensions 
(\ref{TwoTimeResponseTLCFTYlog2}) 
	\begin{align}  
		& C(s, y) = s^{-\frac{d+2}{2}-\nu} 
		y^{-\frac{\alpha \mathcal M}{2}-\frac{d+2}{2}-\nu} 
		\left(1 - \frac{1}{y} \right)^{-\frac{d+2}{2}-\nu} \bigg\{ 1 + \cdots \bigg\} \;,
	\end{align}
where $\cdots$ represent various contributions from the log and log$^2$ extensions. 
From the systematic analysis, we have found that 
our two-time response functions reveal a power-law scaling behavior at early time region,
which is distinct from another power-law scaling at late time region. 
This can be explicitly checked in the figure \ref{fig:NegativeNus}.  
The early time power scaling behaviors are governed by the scaling dimensions of 
the dual field theory operators. In particular, their growth and aging is determined by 
the sign of the parameter $L^2 m^2$. 
The late time behaviors are modified by $\alpha \mathcal M $, the aging parameter. 
If $\alpha =0$, the initial behaviors persist without change, which is expected 
due to its time translation invariance.  
The turning point between these two time regions is marked by the waiting time $s=t_1$. 
The log and log$^2$ corrections provide further modifications that would match 
detailed data by turning available parameters.   

Let us conclude with some observations and future directions toward holographic 
realizations of KPZ universality class. 
Our generalizations of the holographic response functions to general $z$ and $d$ open up some 
possibilities to have contact with the higher dimensional growth and aging phenomena. 
We have done the first attempt to do so in \S \ref{sec:ALCFCriticalExponent}.  
We make some contacts with Kim-Kosterlitz model \cite{KimKosterlitz1989} at higher 
spatial dimensions with an assumptions (\ref{azRelationKPZ}) for some particular dual 
scalar operators. Although it is not perfect, we consider this as a promising sign for 
the future developments along the line. 

We mention two pressing questions we would like to answer in a near future. 
Our holographic model is an infinite system, and thus obtaining the roughness exponent 
``$\mathrm a$'' is rather challenging. Progresses on this point will provide a big step 
toward realizing holographic KPZ class. Assumption (\ref{azRelationKPZ}) is well understood 
in the field theoretical models \cite{BSBook}. 
There Galilean invariance was a crucial ingredient, which is also important in our holographic model. 
Verifying this relation would be an important future challenge.

\section*{Acknowledgments}

We thank to E. Kiritsis, Y. S. Myung, V. Niarchos for discussions and valuable comments 
on the higher dimensional KPZ class and the logarithmic extensions of CFT. 
SH is supported in part by the National Research Foundation of Korea(NRF) grant funded 
by the Korea government(MEST) with the grant number 2012046278. 
SH and JJ are supported by the National Research Foundation of Korea (NRF) grant funded 
by the Korea government(MEST) through the Center for Quantum Spacetime(CQUeST) of Sogang University 
with grant number 2005-0049409. 
JJ is supported in part by the National Research Foundation of Korea(NRF) grant funded 
by the Korea government(MEST) with the grant number 2010-0008359. 
BSK is grateful to the members of the Crete Center for Theoretical Physics, especially 
E. Kiritsis, for his warm hospitality during his visit. 
BSK is supported in part by the Israel Science Foundation (grant number 1468/06).

\end{document}